\pdfoutput=1

\RequirePackage{fix-cm}

\documentclass[twocolumn]{svjour3}

\smartqed

\usepackage{graphicx,amssymb}
\usepackage{epsf,psfig}
\usepackage{txfonts}
\usepackage{natbib}
\usepackage[unicode]{hyperref}

\journalname{Nonlinear Dynamics}

\begin{document}

\title{Escapes in Hamiltonian systems with multiple exit channels -- Part II.}

\author{Euaggelos E. Zotos}

\institute{Department of Physics, School of Science, \\
Aristotle University of Thessaloniki, \\
GR-541 24, Thessaloniki, Greece \\
Corresponding author's email: {evzotos@physics.auth.gr}
}

\date{Received: 9 January 2015 / Accepted: 12 May 2015 / Published online: 29 May 2015}

\titlerunning{Escapes in Hamiltonian systems with multiple exit channels -- Part II.}

\authorrunning{Euaggelos E. Zotos}

\maketitle

\begin{abstract}

We explore the escape dynamics in open Hamiltonian systems with multiple channels of escape continuing the work initiated in Part I. A thorough numerical investigation is conducted distinguishing between trapped (ordered and chaotic) and escaping orbits. The determination of the location of the basins of escape towards the different escape channels and their correlations with the corresponding escape periods of the orbits is undoubtedly an issue of paramount importance. We consider four different cases depending on the perturbation function which controls the number of escape channels on the configuration space. In every case, we computed extensive samples of orbits in both the configuration and the phase space by numerically integrating the equations of motion as well as the variational equations. It was found that in all examined cases regions of non-escaping motion coexist with several basins of escape. The larger escape periods have been measured for orbits with initial conditions in the vicinity of the fractal structure, while the lowest escape rates belong to orbits with initial conditions inside the basins of escape. In addition, we related the model potential with applications in the field of reactive multichannel scattering. We hope that our numerical analysis will be useful for a further understanding of the escape mechanism of orbits in open Hamiltonian systems with two degrees of freedom.

\keywords{Hamiltonian systems; harmonic oscillators; numerical simulations; escapes; fractals}

\end{abstract}

\section{Introduction}
\label{intro}

Over the last decades a huge amount of research work has been devoted on the subject of escaping particles from open dynamical systems. Especially the issue of escape in Hamiltonian systems is a classical problem in nonlinear dynamics (e.g., [\citealp{C90}, \citealp{CE04} -- \citealp{CHLG12}, \citealp{STN02}]). It is well known that several types of Hamiltonian systems have a finite energy of escape. For values of energy lower than the escape energy the equipotential surfaces of the systems are close which means that orbits are bound and therefore escape is impossible. For energy levels above the escape energy on the other hand, the equipotential surfaces open and exit channels emerge through which the particles can escape to infinity. The literature is replete with studies of such ``open" Hamiltonian systems (e.g., [\citealp{BBS09}, \citealp{KSCD99}, \citealp{NH01}, \citealp{STN02}, \citealp{SCK95} -- \citealp{SKCD96}, \citealp{Z13} -- \citealp{Z14b}]). At this point we should emphasize that all the above-mentioned references on escapes in Hamiltonian system are exemplary rather than exhaustive, taking into account that a vast quantity of related literature exists.

Nevertheless, the issue of escaping orbits in Hamiltonian systems is by far less explored than the closely related problem of chaotic scattering. In this situation, a test particle coming from infinity approaches and then scatters off a complex potential. This phenomenon is well investigated as well interpreted from the viewpoint of chaos theory (e.g., [\citealp{BTS96} -- \citealp{BM92}, \citealp{CDO90}, \citealp{CPR75}, \citealp{DGO90} -- \citealp{EJ86}, \citealp{GR89}, \citealp{H88}, \citealp{JRS92} -- \citealp{JT91}, \citealp{LMG00} -- \citealp{ML02}, \citealp{OT93}, \citealp{PH86}, \citealp{RJ94}, \citealp{SASL06} -- \citealp{SS10}]).

During the last half century, dynamical systems made up of perturbed harmonic oscillators have been extensively used in order to describe local motion (i.e., near an equilibrium point) (e.g., [\citealp{AEFR06}, \citealp{CZ12}, \citealp{FLP98a} -- \citealp{HH64}, \citealp{SI79}, \citealp{Z12b} -- \citealp{Z14b}]). In an attempt to reveal and understand the nature of orbits in these systems, scientists have used either numerical (e.g., [\citealp{CZ12}, \citealp{KV08}, \citealp{ZC12}]) or analytical methods (e.g., [\citealp{D91}, \citealp{DE91}, \citealp{E00}, \citealp{ED99}]). Furthermore, potentials made up of harmonic oscillators are frequently used in galactic Astronomy, as a first step for distinguishing between ordered and chaotic local motion in galaxies, since it is widely accepted that the motion of stars near the central region of a galaxy can be approximated by harmonic oscillations (e.g., [\citealp{Z12a}]). One of the most characteristic Hamiltonian systems of two degrees of freedom with three escape channels is undoubtedly the well-known H\'{e}non-Heiles system [\citealp{HH64}]. A huge load of research on the escape properties of this system has been conducted over the years (e.g., [\citealp{AVS01} -- \citealp{AVS03}, \citealp{BBS08}, \citealp{BBS09}, \citealp{FLP98a}, \citealp{FLP98b}]).

In open Hamiltonian systems an issue of paramount importance is the determination of the basins of escape, similar to basins of attraction in dissipative systems or even the Newton-Raphson fractal structures. An escape basin is defined as a local set of initial conditions of orbits for which the test particles escape through a certain exit in the equipotential surface for energies above the escape value. Basins of escape have been studied in many earlier papers (e.g., [\citealp{BGOB88}, \citealp{C02}, \citealp{KY91}, \citealp{PCOG96}, \citealp{SO00}]). The reader can find more details regarding basins of escape in [\citealp{C02}], while the review [\citealp{Z14a}] provides information about the escape properties of orbits in a multi-channel dynamical system of a two-dimensional perturbed harmonic oscillator. The boundaries of an escape basins may be fractal (e.g., [\citealp{AVS09}, \citealp{BGOB88}, \citealp{dML99}]) or even respect the more restrictive Wada property (e.g., [\citealp{AVS01}]), in the case where three or more escape channels coexist in the equipotential surface.

The layout of the present paper is as follows: a detailed presentation of the properties of the Hamiltonian system is given in Section \ref{modpot}. In Section \ref{rescat} we relate our model potential with applications in the field of reactive multichannel scattering. All the computational techniques used in order to determine the character (ordered vs. chaotic and trapped vs. escaping) of orbits are described in Section \ref{cometh}. In the following Section \ref{numres} a thorough numerical analysis of several cases regarding the total number of escape channels is conducted. Our paper ends with Section \ref{disc}, where the discussion and the main conclusions of this work are presented. The text structure of the paper as well as all the numerical methods are the same as in Part I.

\section{Presentation of the Hamiltonian system}
\label{modpot}

The potential of a two-dimensional perturbed harmonic oscillator is
\begin{equation}
V(x,y) = \frac{1}{2}\left(\omega_1^2 x^2 + \omega_2^2 y^2 \right) + \varepsilon V_1(x,y),
\label{genform}
\end{equation}
where $\omega_1$ and $\omega_2$ are the unperturbed frequencies of oscillations along the $x$ and $y$ axes respectively, $\varepsilon$ is the perturbation parameter, while $V_1$ is the function containing the perturbing terms.

As in Part I, we shall use a two-dimensional perturbed harmonic oscillator at the 1:1 resonance, that is when $\omega_1 = \omega_2 = \omega$. Therefore the corresponding potential is
\begin{equation}
V(x,y) = \frac{\omega^2}{2}\left(x^2 + y^2 \right) + \varepsilon V_1(x,y),
\label{pot}
\end{equation}
being $\omega$ the common frequency of oscillations along the two axes. Without the loss of generality, we may set $\omega = 1$ and $\varepsilon = 1$ for more convenient numerical computations.

The basic equations of motion for a test particle with a unit mass $(m = 1)$ are
\begin{equation}
\ddot{x} = - \frac{\partial V}{\partial x}, \ \ \
\ddot{y} = - \frac{\partial V}{\partial y},
\label{eqmot}
\end{equation}
where, as usual, the dot indicates derivative with respect to the time. Furthermore, the variational equations governing the evolution of a deviation vector $\vec{w} = (\delta x, \delta y, \delta \dot{x}, \delta \dot{y})$, which joins the corresponding phase space points of two initially nearby orbits, needed for the calculation of standard chaos indicators (the SALI in our case) are given by
\begin{eqnarray}
\dot{(\delta x)} &=& \delta \dot{x}, \ \ \ \dot{(\delta y)} = \delta \dot{y}, \nonumber \\
(\dot{\delta \dot{x}}) &=& -\frac{\partial^2 V}{\partial x^2}\delta x - \frac{\partial^2 V}{\partial x \partial y}\delta y, \nonumber \\
(\dot{\delta \dot{y}}) &=& -\frac{\partial^2 V}{\partial y \partial x}\delta x - \frac{\partial^2 V}{\partial y^2}\delta y.
\label{variac}
\end{eqnarray}

Consequently, the Hamiltonian to potential (\ref{pot}) (with $\omega = \varepsilon = 1$) reads
\begin{equation}
H = \frac{1}{2}\left(\dot{x}^2 + \dot{y}^2 + x^2 + y^2\right) + V_1(x,y) = h,
\label{ham}
\end{equation}
where $\dot{x}$ and $\dot{y}$ are the momenta per unit mass conjugate to $x$ and $y$ respectively, while $h > 0$ is the numerical value of the Hamiltonian, which is conserved. Thus, an orbit with a given value for it's energy integral is restricted in its motion to regions in which $h \leq V(x,y)$, while all other regions are forbidden to the test particle. The Hamiltonian can also be written in the form
\begin{equation}
H = H_0 + H_1,
\label{ham2}
\end{equation}
with $H_0$ being the integrable term and $H_1$ the non-integrable correction.

The function with the perturbation term $V_1(x,y)$ plays a key role as it determines the location as well as the total number of the escape channels in the configuration $(x,y)$ space. In Part I, we considered perturbing terms that create between two and four escape channels, while now we will investigate the escape dynamics of orbits in the cases where five, six, seven and eight exits are present in the configuration space. At this point we should emphasize that this is the first time that the escape properties of test particles in Hamiltonian systems with more than four escape channels are systematically explored. In order to obtain the appropriate perturbing terms for the required number of escape channels we need a generating function. In polar $(r,\theta)$ coordinates we can easily define functions of the form $r^n \ \sin(n \ \theta)$, where $n$ is the desired number of exits (see e.g., [\citealp{HM09}]). Then we can convert them to rectangular cartesian $(x,y)$ coordinated by following a simple three steps procedure: (i) first split up sums and integer multiples that appear in arguments of trigonometric functions, (ii) expand out products of trigonometric functions into sums of powers, using trigonometric identities when possible and (iii) replace everywhere with $\cos(\theta) = y/r$, $\sin(\theta) = x/r$ and $r = \sqrt{x^2 + y^2}$. In our study we want to work on the $(x,\dot{x})$ phase plane and this type of plane is constructible only if the $V_1(x,y)$ function has terms with even powers regarding the $y$ coordinate. The above-mentioned generating function however, gives terms with even powers of $y$ only for odd values of $n$. Therefore, we need two types of generating functions
\begin{equation}
V_1(r,\theta) = \left\{
\begin{array}{rl}
  r^n \ \sin(n \ \theta), &\mbox{  when $n$ is odd and $n \geq 3,$} \\
  r^n \ \cos(n \ \theta), &\mbox{  when $n$ is even and $n \geq 4,$}
\end{array} \right.
\label{gens}
\end{equation}
regarding how many channels we want the configuration $(x,y)$ space to have. In Appendix A we provide a list of the perturbation functions for the first nine cases, that is for $n \in [2, 10]$\footnote{The perturbation function for the case $n = 2$ is not derived by the generating function.}.

\section{Applications to reactive scattering}
\label{rescat}

Dynamical models with many exits of the form (\ref{pot}) have the nice interpretation as scattering models for rearrangement scattering. We may interpret each exit as a different arrangement in nuclear scattering or molecular scattering. Each channel means a different grouping of particles or atoms into the various fragments. By trajectories entering through one channel and leaving through another channel one can describes nuclear reactions or chemical reactions. As J.R. Taylor writes on page 318 in his well known book [\citealp{T76}], we can imagine multichannel scattering as an irrigation system where water comes in through one channel and goes out through various other channels.

Let us try to explain the basic idea of potential models for rearrangement scattering for the simplest possible case. It is the case of collinear scattering of three particles (e.g., [\citealp{BS90}, \citealp{E88}, \citealp{J91}]). So we have a one dimensional position space and the particles called $A$, $B$, $C$ moving in this position space with coordinates $q_A$, $q_B$, $q_C$. Now we change to relative coordinates $x = q_A - q_B$ and $y = q_C - q_B$. Then the configuration $(x,y)$ plane is the relevant configuration space for all the reactions. The motion of the center of mass of the whole system is irrelevant. Now we imagine a potential with a deep well around the origin and channels along the lines $x = 0$, $y = 0$ and $x - y = 0$. The potential goes to zero rapidly for all other directions. Far away from the origin the channels are straight and of constant depth. First let us assume a negative total energy $E$. Then the trajectory always stays in the central well and the channels. The motion in the central well describes motion where all 3 particles are close and interact. The motion in the channels describes motion of the various asymptotic arrangements. For example think of the channel along $x = 0$ which we call the arrangement channel $C$. In this channel the particles $A$ and $B$ are close enough and interact, while the particle $C$ is far away and moves freely. The general trajectory in channel $C$ moves along the channel with constant velocity in longitudinal direction and at the same time oscillates in transverse direction in the channel potential (i.e., the relative coordinate $q_A - q_B$ oscillates).

We have some kinetic energy $E_k$ for the longitudinal motion and some energy $E_v$ of transversal motion relative to the minimum of the potential channel. The interpretation is like this: A bound state (molecule) of particles (atoms) $A$ and $B$ moves far away from the free particle (atom) $C$. And the molecule $AB$ is in an vibrational state with energy $E_v$. $E_k$ on the other hand, is the kinetic energy of the motion of atom $C$ relative to the bound fragment $AB$. Similar for the other channels. We call the channel along the line $y = 0$ the arrangement channel $A$ and the potential channel along the line $x = y$ the arrangement channel $B$. Now let us assume we find a trajectory which comes in along channel $C$ with transverse energy $E_{v1}$ enters the central potential well, performs complicated motion in the central well for a finite time and then leaves along channel $B$ with transverse energy $E_{v2}$. In position space this event looks like this: Atom $C$ moves in direction of the molecule $AB$ which is in the vibrational state $E_{v1}$. Then they collide, the atoms perform complicated motion for a while and finally the atom $B$ flies away and leaves behind the molecule $AC$ in vibrational state $E_{v2}$. This is the microscopic description of a chemical reaction $AB + C \rightarrow B + AC$.

For total energy larger than 0 there is the additional possibility that all atoms fly away separately. This is the breakup channel 0. The trajectory in the configuration space then leaves the channels and goes away in a direction where the potential is 0. For more particles and for particles in a higher dimensional position space the basic idea remains the same. Take as configuration space a high dimensional space of appropriate relative coordinates and construct a potential with some deep potential well in the middle and some kind of tubes or layers going out into various directions along which certain relative coordinates between particles remain small. Of course, in a 3 dimensional (3D) position space we also include the various rotational states of the bound molecules.

In the potential (\ref{pot}) we could imagine that far away from the origin the total potential converges to some form having the correct asymptotic properties of scattering theory (i.e., the potential goes to zero in most directions and otherwise has a finite number of straight channels of constant depth); such a modification should be doable somehow, while the inside well remain how it is.

\section{Computational methods}
\label{cometh}

In Hamiltonian systems the configuration as well as the phase space is divided into the escaping and non-escaping (trapped) regions. Usually, the vast majority of the trapped space is occupied by initial conditions of regular orbits forming stability islands where a third adelphic integral of motion is present. In many systems however, as we also seen in Part I, trapped chaotic orbits have also been observed. Therefore, we decided to distinguish between regular and chaotic trapped orbits. Over the years, several chaos indicators have been developed in order to determine the character of orbits. In our case, we chose to use the Smaller ALingment Index (SALI) method. The SALI [\citealp{S01}] has been proved a very fast, reliable and effective tool, which is defined as
\begin{equation}
\rm SALI(t) \equiv min(d_-, d_+),
\label{sali}
\end{equation}
where $d_- \equiv \| {\bf{w_1}}(t) - {\bf{w_2}}(t) \|$ and $d_+ \equiv \| {\bf{w_1}}(t) + {\bf{w_2}}(t) \|$ are the alignments indices, while ${\bf{w_1}}(t)$ and ${\bf{w_2}}(t)$, are two deviations vectors which initially point in two random directions. For distinguishing between ordered and chaotic motion, all we have to do is to compute the SALI along time interval $t_{max}$ of numerical integration. In particular, we track simultaneously the time-evolution of the main orbit itself as well as the two deviation vectors ${\bf{w_1}}(t)$ and ${\bf{w_2}}(t)$ in order to compute the SALI. The variational equations (\ref{variac}), as usual, are used for the evolution and computation of the deviation vectors. The time-evolution of SALI strongly depends on the nature of the computed orbit since when the orbit is regular the SALI exhibits small fluctuations around non zero values, while on the other hand, in the case of chaotic orbits the SALI after a small transient period it tends exponentially to zero approaching the limit of the accuracy of the computer $(10^{-16})$. Therefore, the particular time-evolution of the SALI allow us to distinguish fast and safely between regular and chaotic motion (e.g., [\citealp{ZC13}]). Nevertheless, we have to define a specific numerical threshold value for determining the transition from regularity to chaos. After conducting extensive numerical experiments, integrating many sets of orbits, we conclude that a safe threshold value for the SALI is the value $10^{-7}$. In order to decide whether an orbit is regular or chaotic, one may use the usual method according to which we check after a certain and predefined time interval of numerical integration, if the value of SALI has become less than the established threshold value. Therefore, if SALI $\leq 10^{-7}$ the orbit is chaotic, while if SALI $ > 10^{-7}$ the orbit is regular. For the computation of SALI we used the \verb!LP-VI! code [\citealp{CMD14}], a fully operational routine which efficiently computes a suite of many chaos indicators for dynamical systems in any number of dimensions.

For investigating the escape escape process in our Hamiltonian system, we need to define samples of orbits whose nature (escaping or trapped) will be identified. For this purpose we define for each value of the energy (all tested energy levels are always above the escape energy), dense, uniform grids of initial conditions regularly distributed in the area allowed by the value of the energy. Our investigation takes place both in the configuration $(x,y)$ and the phase $(x,\dot{x})$ space for a better understanding of the escape mechanism. In both cases, the step separation of the initial conditions along the axes (or in other words the density of the grids) was controlled in such a way that always there are about 50000 orbits (a maximum grid of 225 $\times$ 225 equally spaced initial conditions of orbits). For each initial condition, we integrated the equations of motion (\ref{eqmot}) as well as the variational equations (\ref{variac}) using a double precision Bulirsch-Stoer \verb!FORTRAN 77! algorithm (e.g., [\citealp{PTVF92}]) with a small time step of order of $10^{-2}$, which is sufficient enough for the desired accuracy of our computations (i.e., our results practically do not change by halving the time step). Our previous experience suggests that the Bulirsch-Stoer integrator is both faster and more accurate than a double precision Runge-Kutta-Fehlberg algorithm of order 7 with Cash-Karp coefficients. In all cases, the energy integral (Eq. (\ref{ham})) was conserved better than one part in $10^{-11}$, although for most orbits it was better than one part in $10^{-12}$.

In Hamiltonian systems with escapes an issue of paramount importance is the determination of the position as well as the time at which an orbit escapes. When the value of the energy $h$ is smaller than the escape energy, the Zero Velocity Curves (ZVCs) are closed. On the other hand, when $h > h_{esc}$ the equipotential curves are open and extend to infinity. An open ZVC consists of several branches forming channels through which an orbit can escape to infinity. At every opening there is a highly unstable periodic orbit close to the line of maximum potential [\citealp{C79}] which is called a Lyapunov orbit. Such an orbit reaches the ZVC, on both sides of the opening and returns along the same path thus, connecting two opposite branches of the ZVC. Lyapunov orbits are very important for the escapes from the system, since if an orbit intersects any one of these orbits with velocity pointing outwards moves always outwards and eventually escapes from the system without any further intersections with the surface of section (see e.g., [\citealp{C90}]). Additional details regarding the escape criteria are given in the Appendix B. The passage of orbits through Lyapunov orbits and their subsequent escape to infinity is the most conspicuous aspect of the transport, but crucial features of the bulk flow, especially at late times, appear to be controlled by diffusion through cantori, which can trap orbits far vary long time periods.

For the numerical integration we set a maximum time equal to $10^5$ time units. Our previous experience in this subject indicates that usually orbits need considerable less time to find one of the exits in the limiting surface and eventually escape from the system (obviously, the numerical integration is effectively ended when an orbit passes through one of the escape channels and intersects one of the unstable Lyapunov orbits). Nevertheless, we decided to use such a vast integration time just to be sure that all orbits have enough time in order to escape. Remember that there are the so called ``sticky orbits" which behave as regular ones and their true chaotic character is revealed only after long time intervals of numerical integration. Here we should clarify that orbits which do not escape after a numerical integration of $10^5$ time units are considered as non-escaping or trapped.

\section{Numerical results}
\label{numres}

The main target is to distinguish between trapped and escaping orbits for values of energy larger than the escape energy where the Zero Velocity Curves are open and several channels of escape are present. Furthermore, two important properties of the orbits will be investigated: (i) the directions or channels through which the particles escape and (ii) the time-scale of the escapes (we shall also use the term escape period). In particular, we examine these aspects for various values of the energy $h$, as well as for four different types of perturbation function $V_1(x,y)$. The grids of initial conditions of orbits whose properties will be determined are defined as follows: for the configuration $(x,y)$ space we consider orbits with initial conditions $(x_0, y_0)$ with $\dot{x_0} = 0$, while the initial value of $\dot{y_0}$ is always obtained from the energy integral of motion (\ref{ham}) as $\dot{y_0} = \dot{y}(x_0,\dot{x_0},h) > 0$. Similarly, for the phase $(x,\dot{x})$ space we consider orbits with initial conditions $(x_0, \dot{x_0})$ with $y_0 = 0$, while again the initial value of $\dot{y_0}$ is obtained from the Hamiltonian (\ref{ham}).

Our numerical calculations indicate that in almost all cases, apart from the escaping orbits there is an amount of non-escaping orbits. In general terms, the majority of non-escaping regions corresponds to initial conditions of regular orbits, where a third integral of motion is present, restricting their accessible phase space and therefore hinders their escape. However, there are also chaotic orbits which do not escape within the predefined time interval and remain trapped for vast periods until they eventually escape to infinity. At this point, it should be emphasized and clarified that these trapped chaotic orbits cannot be considered, by no means, neither as sticky orbits nor as super sticky orbits with sticky periods larger than $10^5$ time units. Sticky orbits are those who behave regularly for long time periods before their true chaotic nature is fully revealed. In our case on the other hand, this type of orbits exhibit chaoticity very quickly as it takes no more than about 100 time units for the SALI to cross the threshold value (SALI $\ll 10^{-7}$), thus identifying beyond any doubt their chaotic character. Therefore, we decided to classify the initial conditions of orbits in both the configuration and phase space into three main categories: (i) orbits that escape through one of the escape channels, (ii) non-escaping regular orbits and (iii) trapped chaotic orbits.

Here we would like to point out that all the following subsections containing the results of the four cases are formed having in mind flexibility. According the current text structure the reader can read any of the four subsections and have a clear view of the properties of the corresponding Hamiltonian system because each subsection is practical text-autonomous.

\subsection{Case I: Five channels of escape}
\label{case1}

\begin{figure*}[!tH]
\centering
\resizebox{0.8\hsize}{!}{\includegraphics{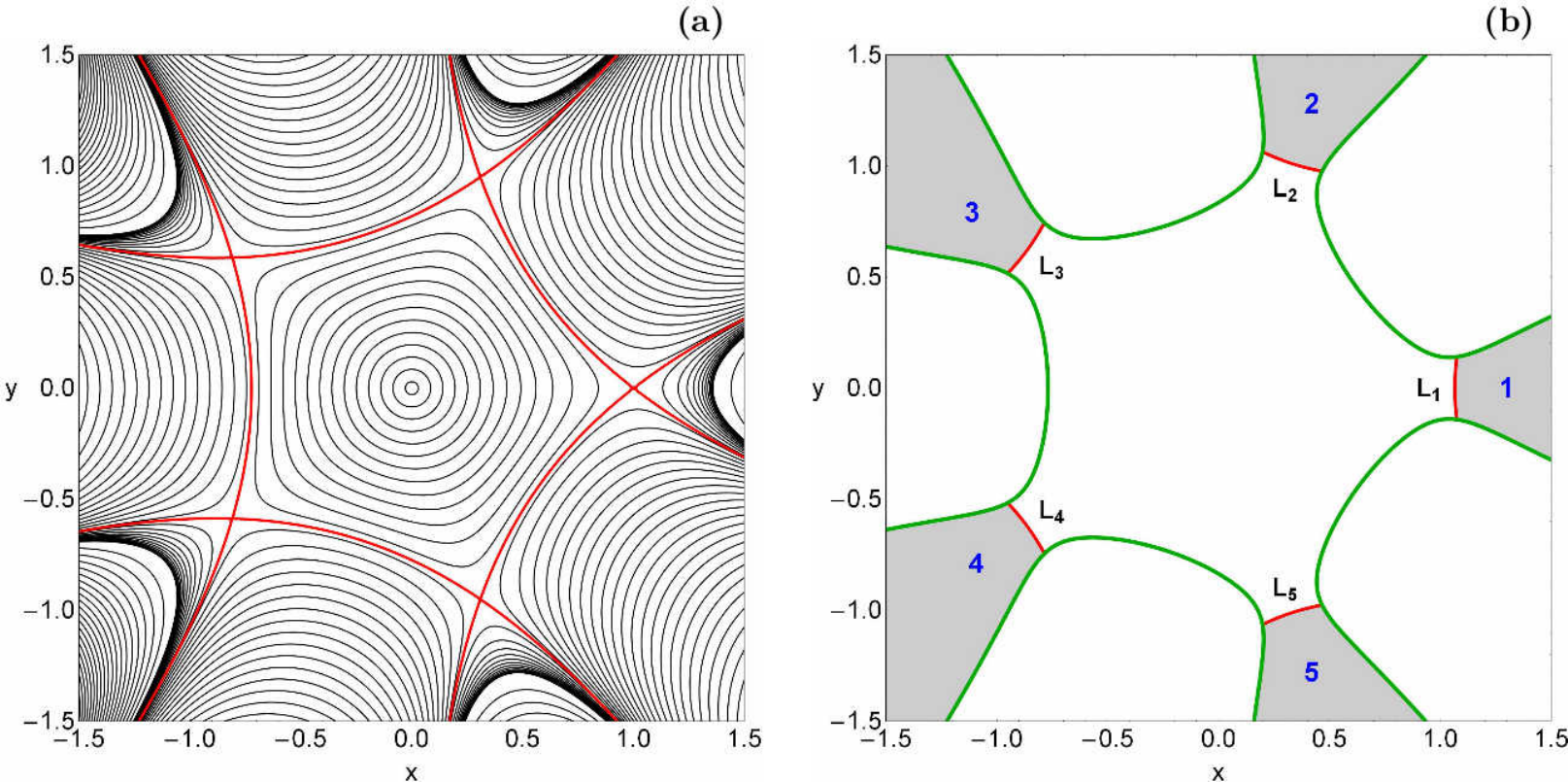}}
\caption{(a-left): Equipotential curves of the total potential (\ref{pot}) for various values of the energy $h$, when five escape channels are present. The equipotential curve corresponding to the energy of escape is shown with red color; (b-right): The open ZVC at the configuration $(x,y)$ plane when $h = 0.35$. $L_i$, $i=1,...,5$ indicate the five unstable Lyapunov orbits plotted in red.}
\label{pot5}
\end{figure*}

In this case $(n = 5)$, the non-integrable part of the Hamiltonian according to the first generating function (\ref{gens}) is
\begin{equation}
H_1 = V_1(x,y) = - \frac{1}{5}\left(x^5 - 10 x^3 y^2 + 5 x y^4 \right),
\label{ham1}
\end{equation}
and the corresponding escape energy equals to 3/10. The total Hamiltonian of the system $H = H_0 + H_1$ has a special symmetry, that is $H$ is symmetric with respect to $y \rightarrow - y$. The equipotential curves of the total potential (\ref{pot}) for various values of the energy $h$ are shown in Fig. \ref{pot5}a. The equipotential corresponding to the energy of escape $h_{esc}$ is plotted with red color in the same plot. The open ZVC at the configuration $(x,y)$ plane when $h = 0.35 > h_{esc}$ is presented with green color in Fig. \ref{pot5}b and the five channels of escape are shown. In the same plot, we denote the five unstable Lyapunov orbits by $L_i$, $i = 1,...,5$ using red color.

\begin{figure*}[!tH]
\centering
\resizebox{\hsize}{!}{\includegraphics{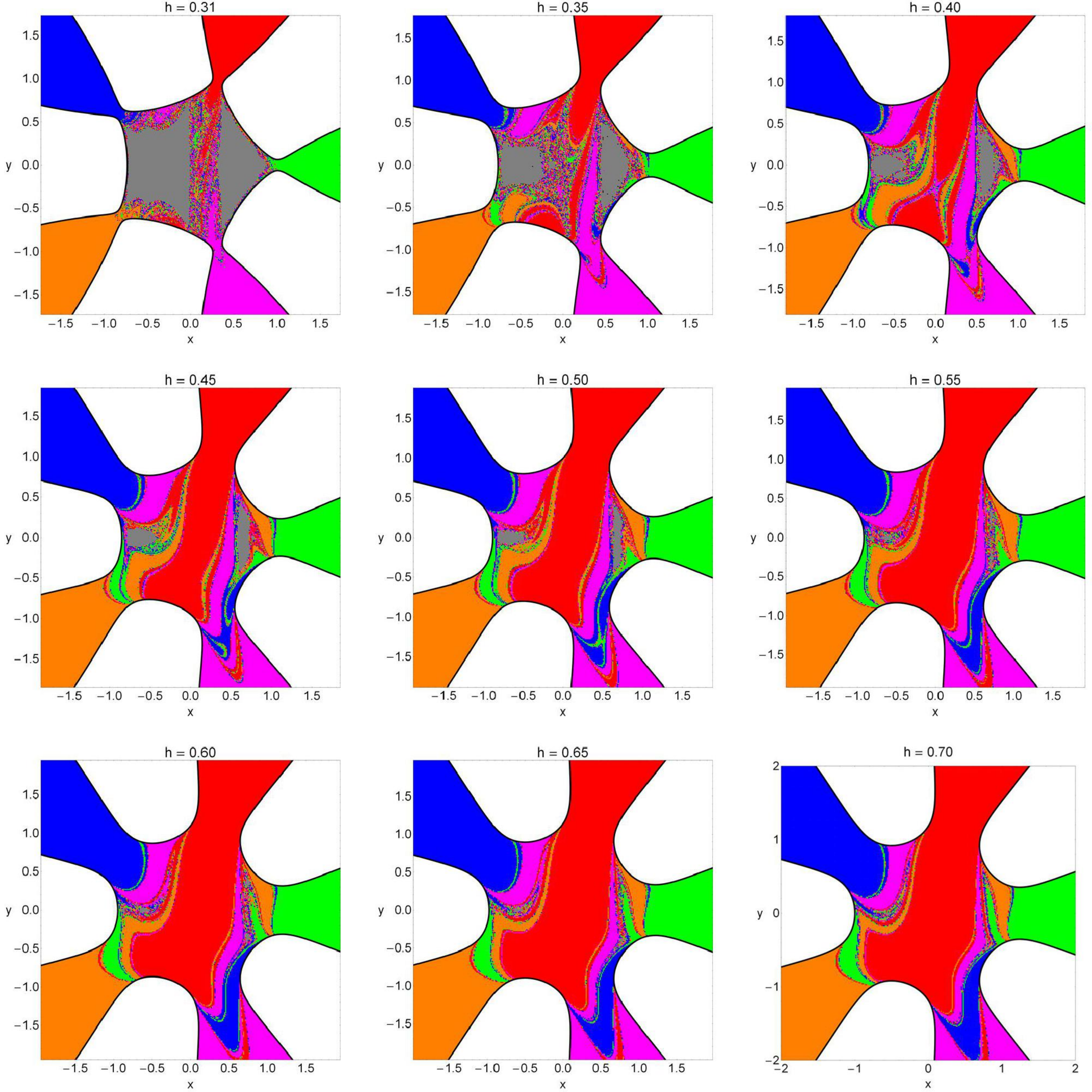}}
\caption{The structure of the configuration $(x,y)$ plane for several values of the energy $h$, distinguishing between different escape channels. The color code is as follows: Escape through channel 1 (green); escape through channel 2 (red); escape through channel 3 (blue); escape through channel 4 (orange); escape through channel 5 (magenta); non-escaping regular (gray); trapped chaotic (black).}
\label{xy5}
\end{figure*}

\begin{figure*}[!tH]
\centering
\resizebox{\hsize}{!}{\includegraphics{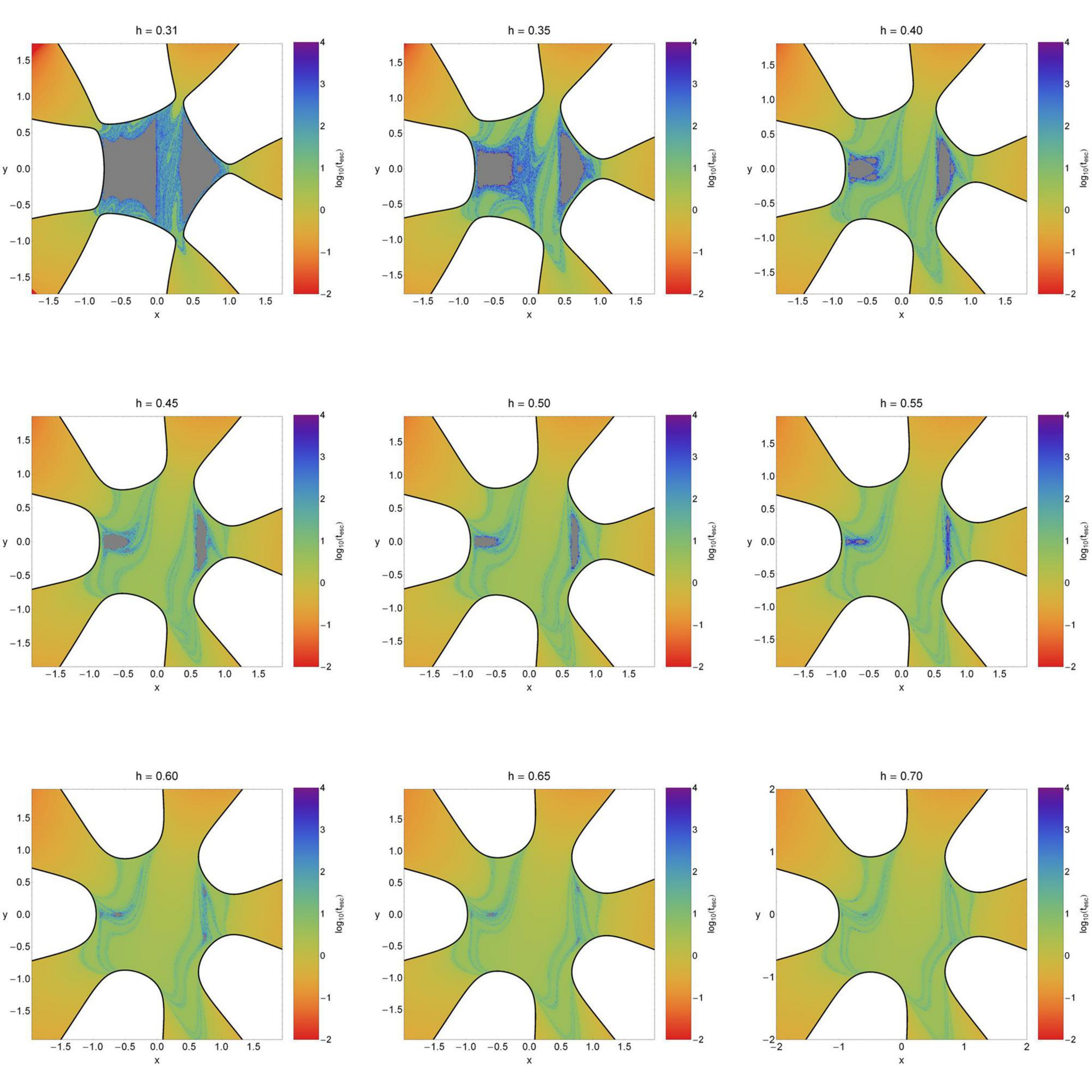}}
\caption{Distribution of the escape times $t_{\rm esc}$ of the orbits on the $(x,y)$ plane. The darker the color, the larger the escape time. Trapped and non-escaping orbits are indicated by gray color.}
\label{xyt5}
\end{figure*}

\begin{figure}
\includegraphics[width=\hsize]{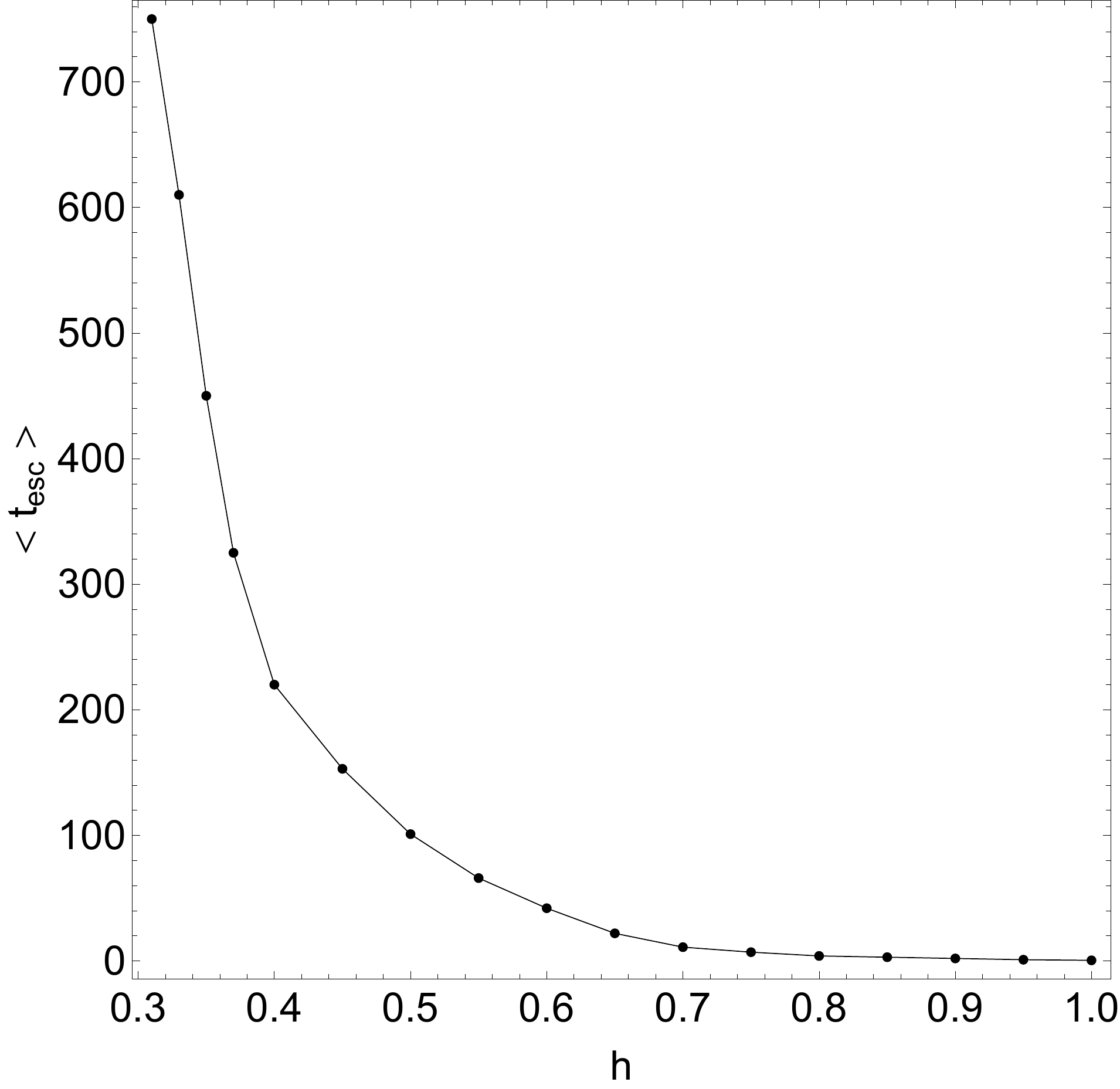}
\caption{Evolution of the average escape time of orbits $< t_{\rm esc} >$ on the configuration $(x,y)$ space as a function of the total orbital energy $h$.}
\label{tesc}
\end{figure}

\begin{figure*}[!tH]
\centering
\resizebox{\hsize}{!}{\includegraphics{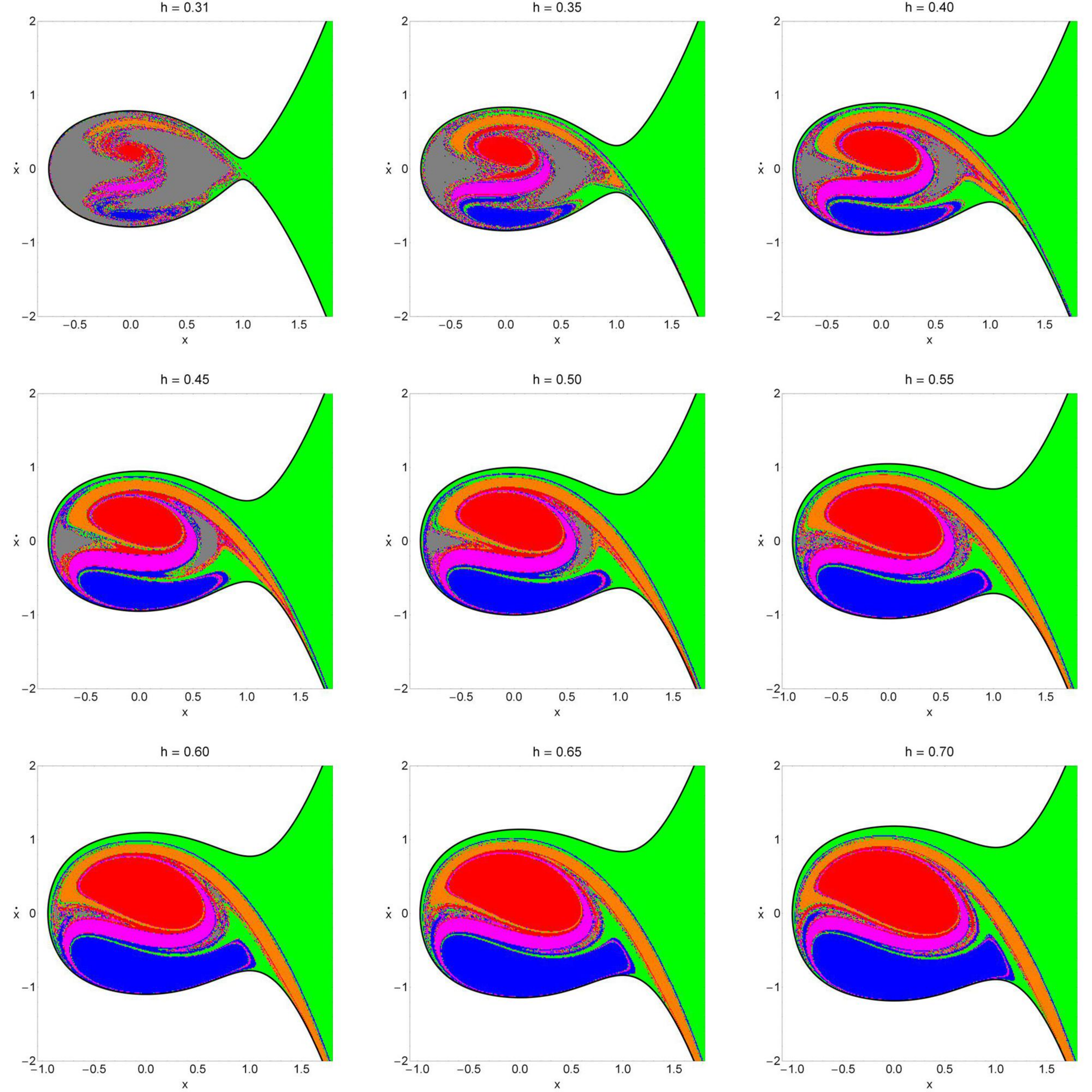}}
\caption{The structure of the phase $(x,\dot{x})$ plane for several values of the energy $h$, distinguishing between different escape channels. The color code is the same as in Fig. \ref{xy5}.}
\label{xpx5}
\end{figure*}

\begin{figure*}[!tH]
\centering
\resizebox{\hsize}{!}{\includegraphics{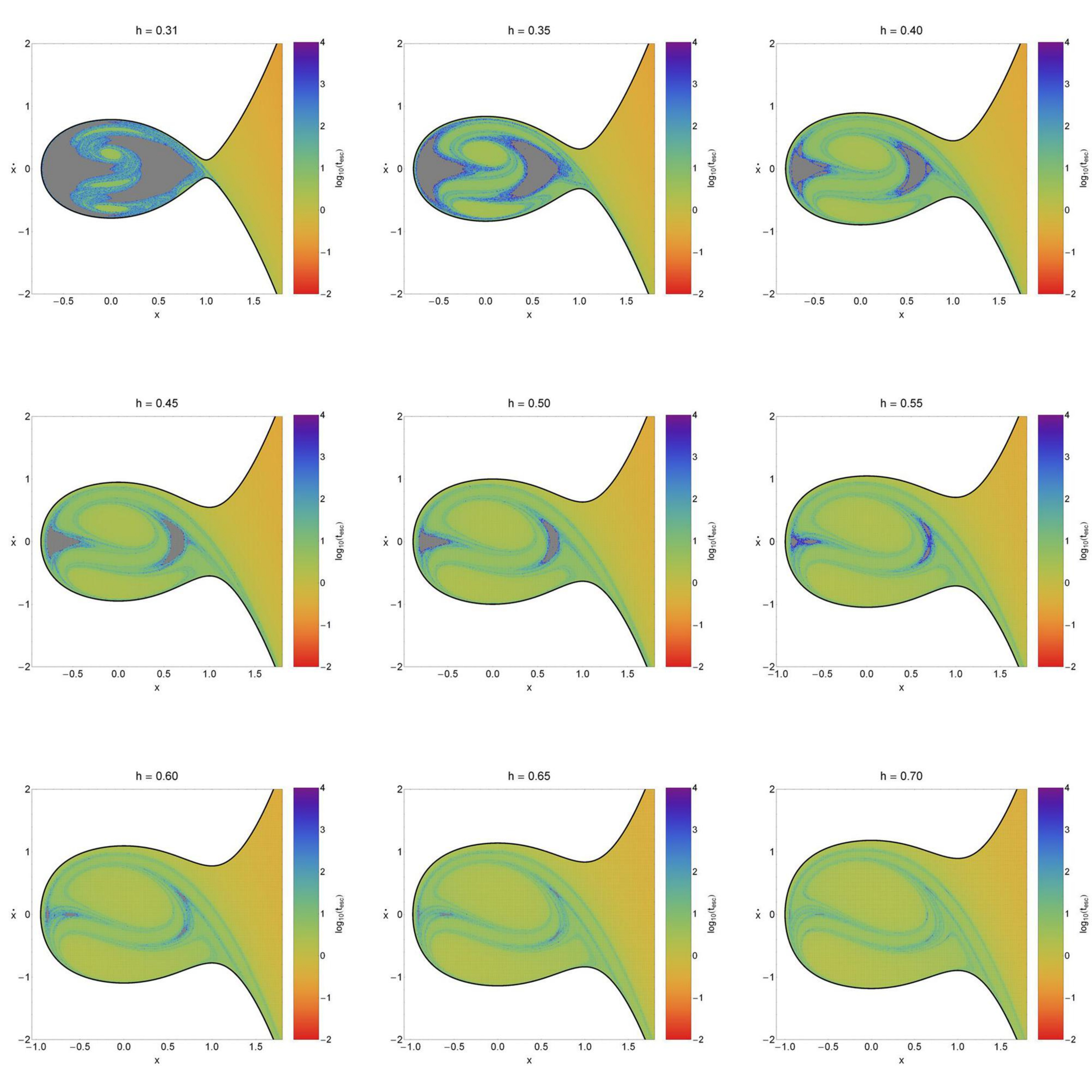}}
\caption{Distribution of the escape times $t_{\rm esc}$ of the orbits on the $(x,\dot{x})$ plane. The darker the color, the larger the escape time. Trapped and non-escaping orbits are indicated by gray color.}
\label{xpxt5}
\end{figure*}

\begin{figure*}[!tH]
\centering
\resizebox{0.8\hsize}{!}{\includegraphics{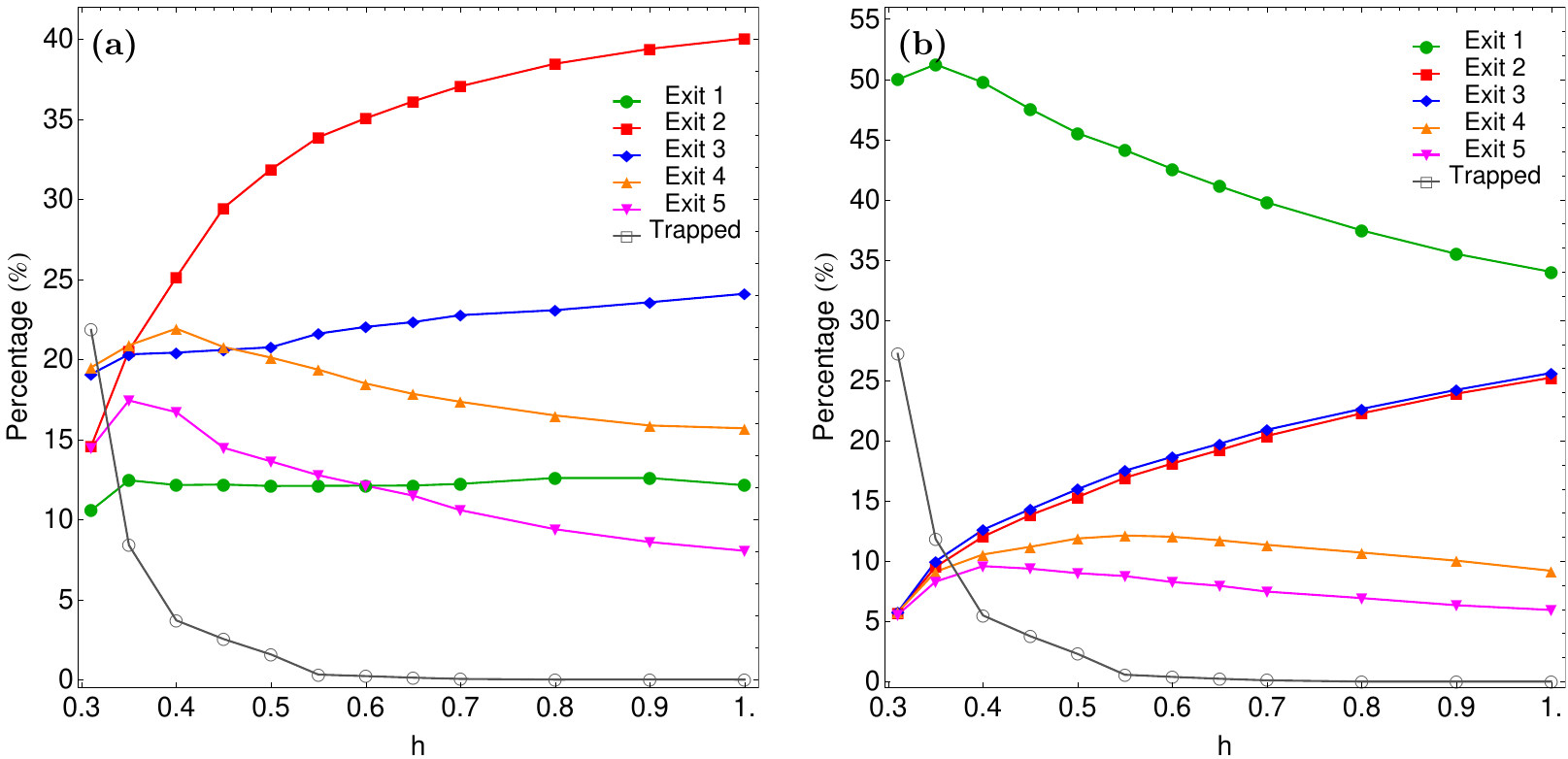}}
\caption{Evolution of the percentages of trapped and escaping orbits when varying the energy $h$ (a-left): on the configuration $(x,y)$ plane and (b-right): on the phase $(x,\dot{x})$ plane.}
\label{percs5}
\end{figure*}

We will investigate the trapped or escape dynamics of test particles for values of energy in the set $h$ = \{0.31, 0.35, 0.40, 0.45, 0.50, 0.55, 0.60, 0.65, 0.70\}. Our exploration begins in the configuration $(x,y)$ space and in Fig. \ref{xy5} we present the structure of the $(x,y)$ plane for several values of the energy. Each initial condition is colored according to the escape channel through which the particular orbit escapes. The gray regions on the other hand, denote initial conditions where the test particles move in regular orbits and do not escape, while trapped chaotic orbits are indicated with black. The outermost solid line is the Zero Velocity Curve (limiting curve) which is defined as $V(x,y) = h$. It is seen that for values of energy larger but yet very close to the escape energy $(h < 0.40)$ a large portion of the $(x,y)$ plane is covered by stability islands which correspond to initial conditions of non-escaping regular orbits which are surrounded by a very rich fractal structure. Moreover, looking carefully the grids we observe that there is a highly sensitive dependence of the escape process on the initial conditions, that is, a slight change in the initial conditions makes the test particle escape through another channel, which is of course a classical indication of chaos. As the value of the energy increases the stability islands and the amount of non-escaping and trapped orbits is reduced and basins of escape emerge. Indeed, when $h = 0.70$ almost all the computed orbits of the grid escape and there is no indication of bounded motion or whatsoever. By the term basin of escape, we refer to a local set of initial conditions that corresponds to a certain escape channel. The escape basins become smoother and more well-defined as the energy increases and the degree of fractality decreases\footnote{The fat-fractal exponent increases, approaching the value 1 which means no fractal geometry, when the energy of the system is high enough (see [\citealp{BBS08}]).}. The fractality is strongly related with the unpredictability in the evolution of a dynamical system. In our case, it can be interpreted that for high enough energy levels, the test particles escape very fast from the scattering region and therefore, the system's predictability increases. It is seen in channel 5 that outside the Lyapunov orbit there is a stream of initial conditions of orbits which even though they are launched outside the unstable the Lyapunov orbits they escape from another exit, meaning that before escape they first enter the interior region.

The distribution of the escape times $t_{\rm esc}$ of orbits on the $(x,y)$ plane is given in the Fig. \ref{xyt5}, where light reddish colors correspond to fast escaping orbits, dark blue/purpe colors indicate large escape periods, while gray color denote both trapped chaotic and non-escaping regular orbits. It is observed that for $h = 0.31$, that is a value of energy just above the escape energy, the escape periods of the majority of orbits are huge corresponding to tens of thousands of time units. This however, is anticipated because in this case the width of the escape channels is very small and therefore, the orbits should spend much time inside the equipotential surface until they find one of the five openings and eventually escape to infinity. As the value of the energy increases however, the escape channels become more and more wide leading to faster escaping orbits, which means that the escape period decreases rapidly. We found that the longest escape rates correspond to initial conditions near the vicinity of the fractal regions. On the other hand, the shortest escape periods have been measured for the regions without sensitive dependence on the initial conditions (basins of escape), that is, those far away from the fractal basin boundaries. We would like to emphasize that by definition, the fractal basin boundaries contain initial conditions of orbits that will never escape from the system, as it coincides with the stable manifold of the non-attracting chaotic set, also known as chaotic saddle or strange saddle, that is formed by a set of Lebesgue measure zero of orbits that will never escape from the scattering region for both $t \rightarrow \infty$ or $t \rightarrow - \infty$. It is known that at the critical energy the escape time is infinity and it decreases if one moves away from the critical value. The evolution of the average value of the escape time $< t_{\rm esc} >$ of orbits on the configuration $(x,y)$ space as a function of the total orbital energy $h$ is given in Fig. \ref{tesc}. It is seen, that for low values of energy, just above the escape value, the average escape time of orbits is about 750 time units, however it reduces rapidly tending asymptotically to zero which refers to orbits that escape almost immediately from the system.

The structure of the phase $(x,\dot{x})$ plane for the same set of values of the energy is shown in Fig. \ref{xpx5}. A similar behavior to that discussed for the configuration $(x,y)$ plane can be seen. The outermost black solid line is the limiting curve which is defined as
\begin{equation}
f(x,\dot{x}) = \frac{1}{2}\dot{x}^2 + V(x, y = 0) = h.
\label{zvc}
\end{equation}
Here we must clarify that this $(x,\dot{x})$ phase plane is not a classical Poincar\'{e} Surface of Section (PSS), simply because escaping orbits in general, do not intersect the $y = 0$ axis after a certain time, thus preventing us from defying a recurrent time. A classical Poincar\'{e} surface of section exists only if orbits intersect an axis, like $y = 0$, at least once within a certain time interval. Nevertheless, in the case of escaping orbits we can still define local surfaces of section which help us to understand the orbital behavior of the dynamical system. It is interesting to note that the limiting curve is open at the right part due to the $x^5$ term entering the perturbation function. In the phase planes of Fig. \ref{xpx5} one can distinguish fractal regions where it is impossible to predict the particular escape channel and regions occupied by escape basins. These basins are either broad well-defined regions, or elongated bands of complicated structure spiralling around the center. Once more we observe that for values of energy close to the escape energy there is a substantial amount of non-escaping regular orbits and the degree of fractalization of the rest phase plane is high. As we proceed to higher energy levels however, the rate of non-escaping regular orbits heavily reduces, the phase plane becomes less and less fractal and is occupied by well-defined basins of escape. We would like to note that at the right open part of the $(x,\dot{x})$ planes there is flow of initial conditions which extends asymptotically to infinity. The distribution of the escape times $t_{\rm esc}$ of orbits on the $(x,\dot{x})$ plane is shown in Fig. \ref{xpxt5}. It is evident that orbits with initial conditions inside the exit basins escape from the system very quickly, or in other words, they possess extremely low escape periods. On the contrary, orbits with initial conditions located in the fractal parts of the phase plane need considerable amount of time in order to escape.

The following Fig. \ref{percs5}a shows the evolution of the percentages of trapped and escaping orbits on the configuration $(x,y)$ plane when the value of the energy $h$ varies. Here we would like to point out that we decided to merge the percentages of non-escaping regular and trapped chaotic orbits together because our computations indicate that always the rate of trapped chaotic orbits is extremely small (less than 1\%) and therefore, it does not contribute to the overall orbital structure of the dynamical system. We observe that when $h = 0.31$, that is just above the escape energy, trapped motion is the most populated family occupying about 22\% of the configuration plane, while escaping orbits through exits 2 and 3 have the same rates with escaping orbits through channels 5 and 4, respectively. As the value of the energy increases however, the rate of trapped orbits drops rapidly and for $h > 0.60$ it practically vanishes. At the same time, the percentage of orbits escaping through exit channel 2 increases steadily and at the highest energy level studied it corresponds to about 40\% of the configuration $(x,y)$ plane. The rates of escaping orbits through channels 4 and 5 exhibit a similar slow reduction for $h > 0.4$, while the percentages of escaping orbits through exits 1 and 3 seem less unaffected by the shifting of the energy being almost unperturbed around 12\% and 23\%, respectively. Therefore, one may conclude that for high energy levels $(h > 0.60)$, all orbits in the configuration $(x,y)$ plane escape and about 40\% of them choose channel 2. In the same vein we present in Fig. \ref{percs5}b the evolution of the percentages of trapped and escaping orbits on the phase plane as a function of the energy $h$. It is observed that the pattern and the evolution of the percentages is completely different with respect to that discussed in Fig. \ref{percs5}a regarding the configuration plane. We see that escaping orbits through exit channel 1 dominate throughout, even though their rate reduces with increasing energy. Moreover, the percentages of escaping orbits through exits 2 and 3 display an identical increase from 5\% to 25\%, while the rates of exits 4 and 5 are much smaller (less than 10\%). The only similarity with the configuration plane is the evolution pattern of trapped orbits. Taking all into account we can deduce that in the configuration space an orbit is more likely to escape form channel 1, while for sufficiently enough values of energy $(h > 1)$, we have numerical evidence that the rates of exits 1, 2 and 3 seem to converge thus sharing about 90\% of the phase space.

\subsection{Case II: Six channels of escape}
\label{case2}

\begin{figure*}[!tH]
\centering
\resizebox{0.8\hsize}{!}{\includegraphics{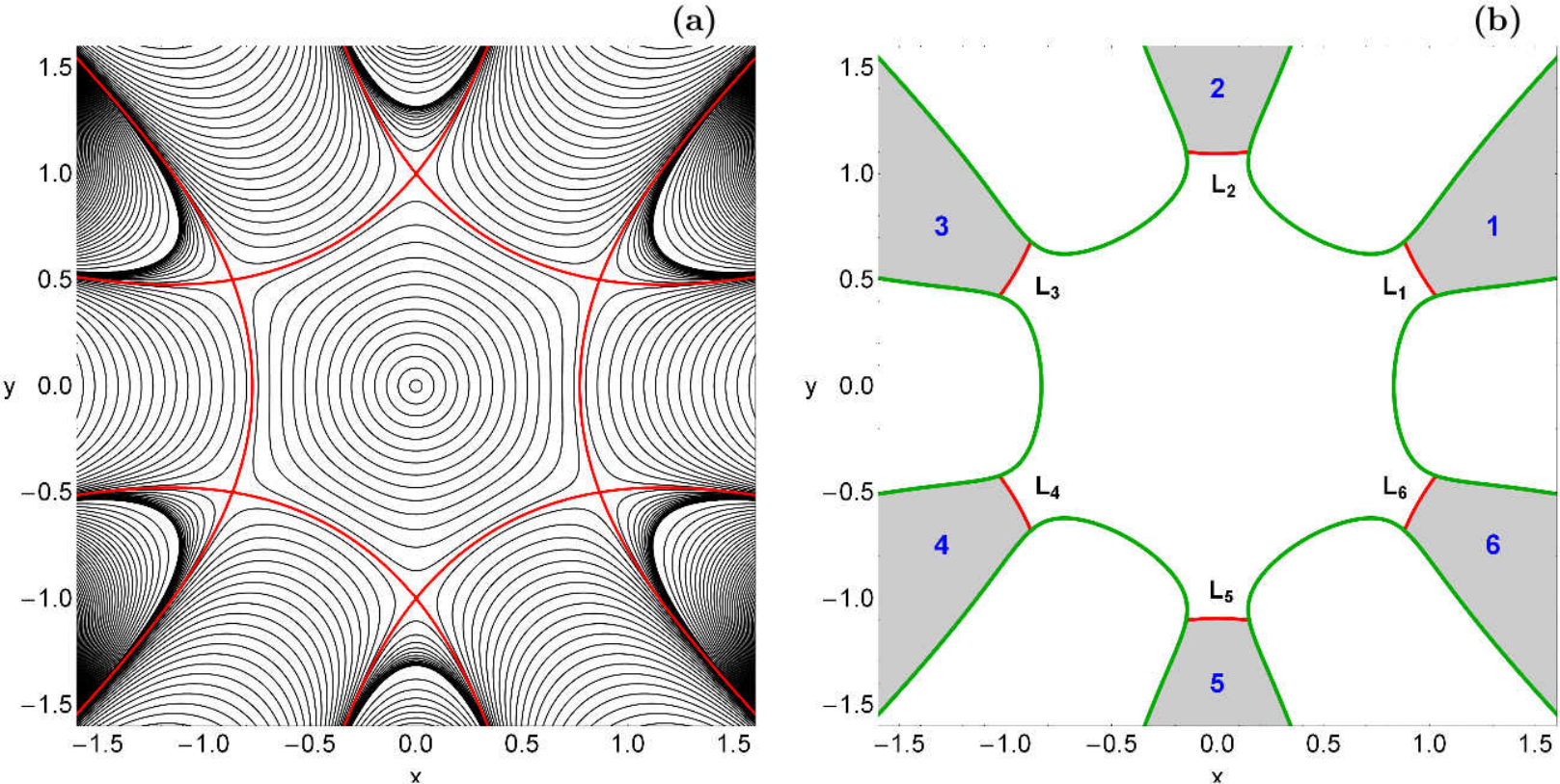}}
\caption{(a-left): Equipotential curves of the total potential (\ref{pot}) for various values of the energy $h$, when six escape channels are present. The equipotential curve corresponding to the energy of escape is shown with red color; (b-right): The open ZVC at the configuration $(x,y)$ plane when $h = 0.40$. $L_i$, $i=1,...,6$ indicate the six unstable Lyapunov orbits plotted in red.}
\label{pot6}
\end{figure*}

We continue our exploration of escapes in a Hamiltonian system with six exit channels with escape energy equal to 1/3. In order to obtain this number of exits $(n = 6)$ in the limiting curve in the configuration $(x,y)$ plane, the perturbation term should be
\begin{equation}
V_1(x,y) = - \frac{1}{6}\left(x^6 + 15 x^4 y^2 - 15 x^2 y^4 + y^6 \right),
\label{ham2}
\end{equation}
according to the second generating function of Eqs. (\ref{gens}). The corresponding Hamiltonian $H = H_0 + H_1$ is invariant under $x \rightarrow - x$ and/or $y \rightarrow - y$. In Fig. \ref{pot6}a we see the equipotential curves of the total potential (\ref{pot}) for various values of the energy $h$, while the equipotential corresponding to the energy of escape $h_{esc}$ is plotted with red color in the same plot. Furthermore, the open ZVC at the configuration $(x,y)$ plane when $h = 0.4 > h_{esc}$ is presented with green color in Fig. \ref{pot6}b and the six channels of escape are shown. In the same figure, the six unstable Lyapunov orbits $L_i$, $i = {1,...,6}$ are denoted using red color.

\begin{figure*}[!tH]
\centering
\resizebox{\hsize}{!}{\includegraphics{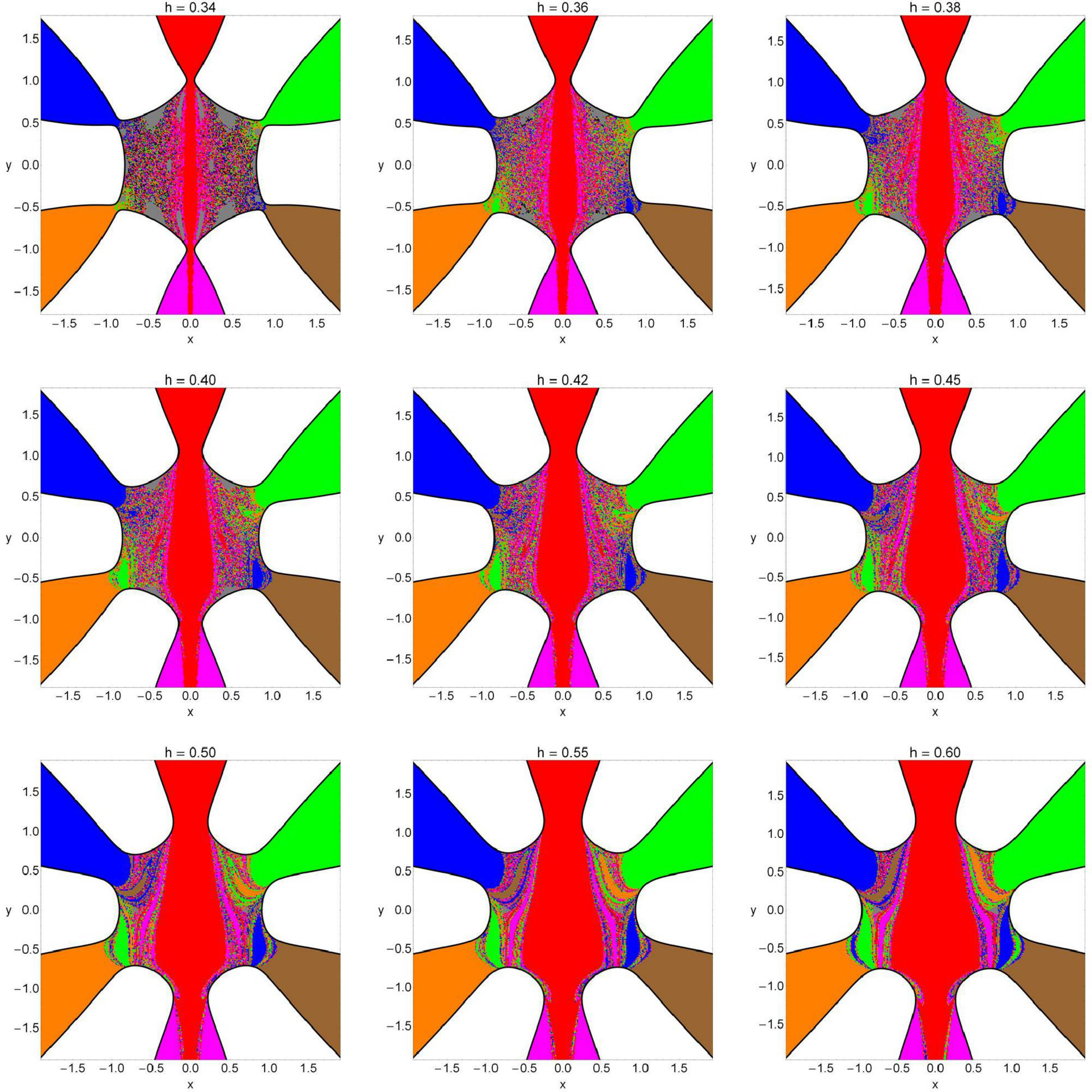}}
\caption{The structure of the configuration $(x,y)$ plane for several values of the energy $h$, distinguishing between different escape channels. The color code is as follows: Escape through channel 1 (green); escape through channel 2 (red); escape through channel 3 (blue); escape through channel 4 (orange); escape through channel 5 (magenta); escape through channel 6 (brown); non-escaping regular (gray); trapped chaotic (black).}
\label{xy6}
\end{figure*}

\begin{figure*}[!tH]
\centering
\resizebox{\hsize}{!}{\includegraphics{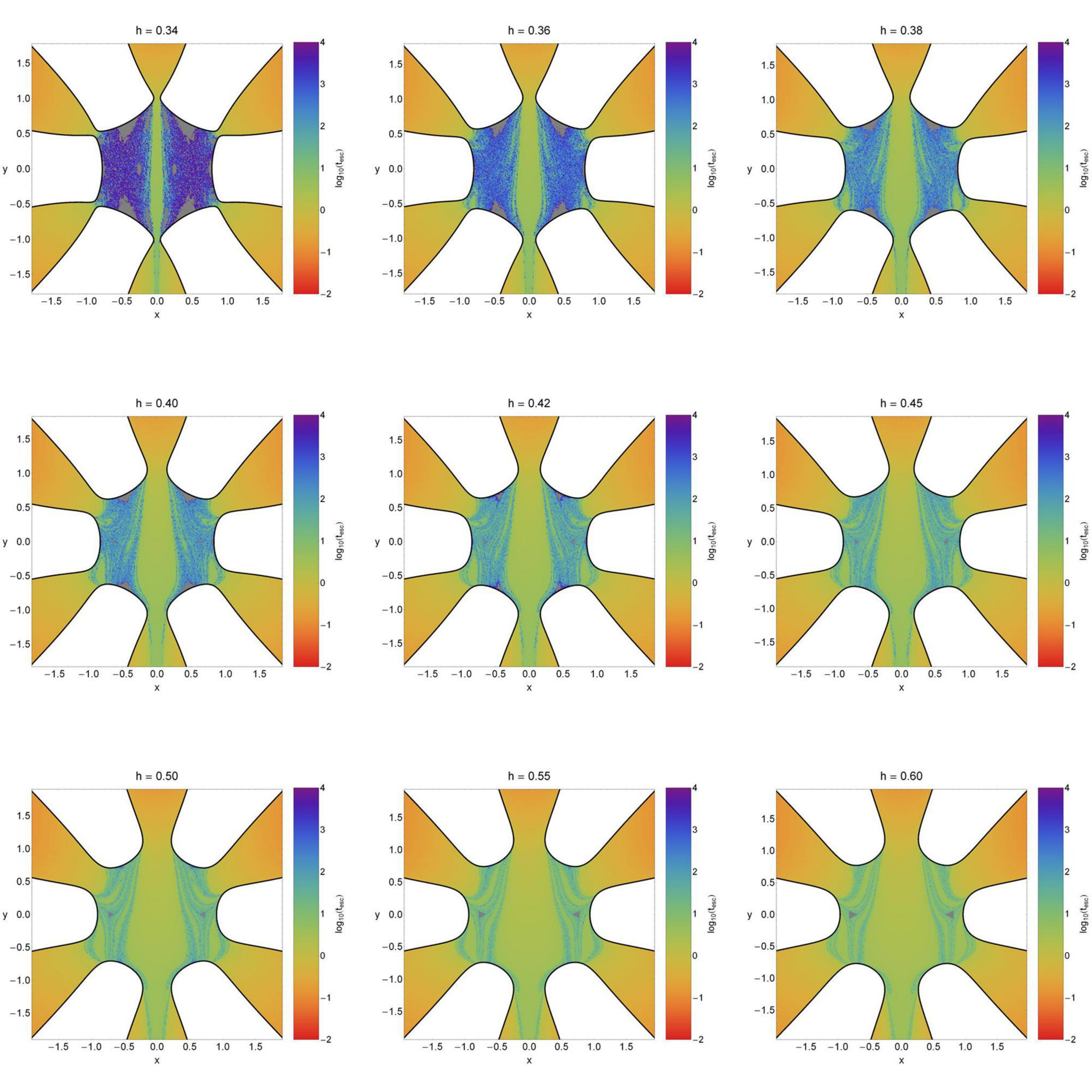}}
\caption{Distribution of the escape times $t_{\rm esc}$ of the orbits on the $(x,y)$ plane. The darker the color, the larger the escape time. Trapped and non-escaping orbits are indicated by gray color.}
\label{xyt6}
\end{figure*}

\begin{figure*}[!tH]
\centering
\resizebox{\hsize}{!}{\includegraphics{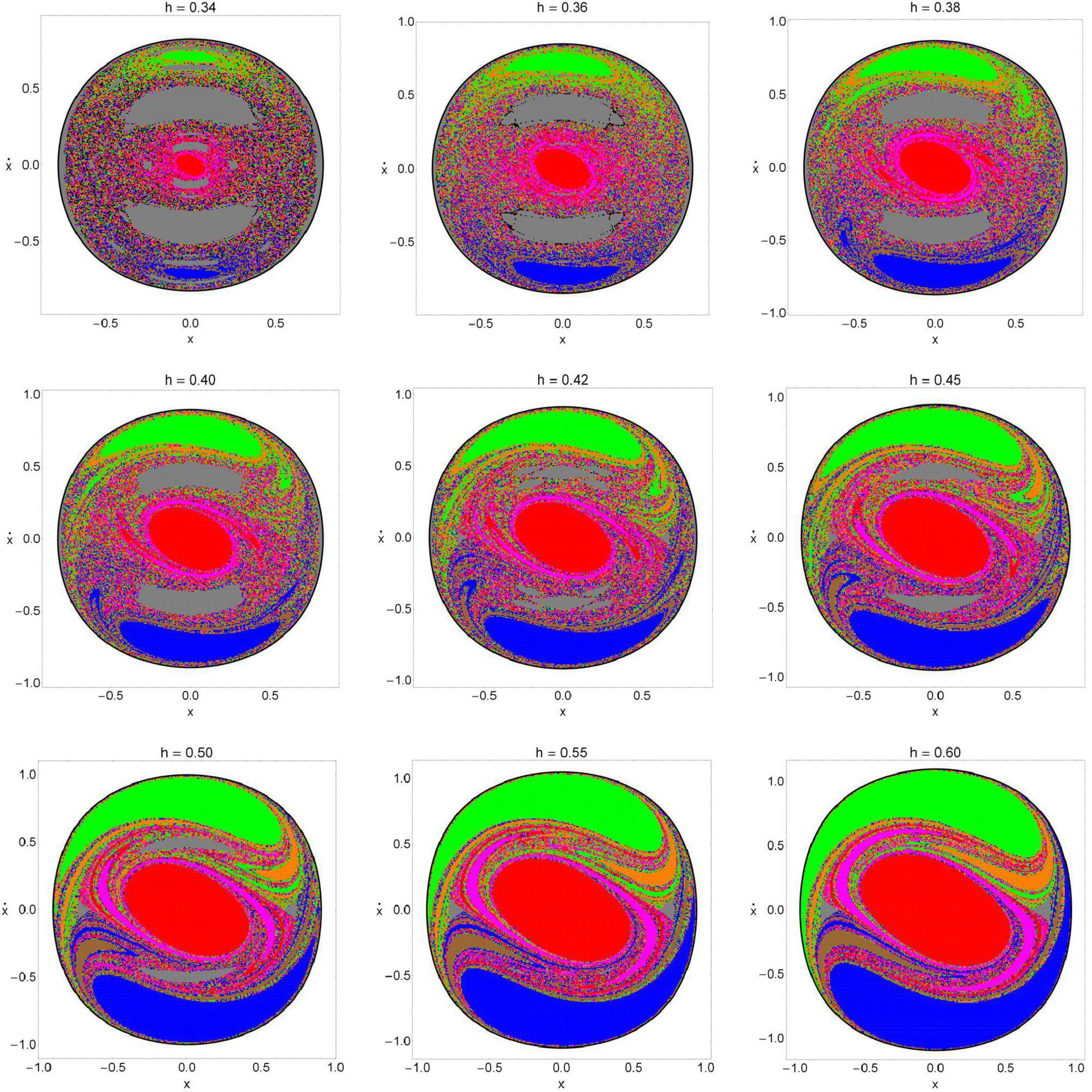}}
\caption{The structure of the phase $(x,\dot{x})$ plane for several values of the energy $h$, distinguishing between different escape channels. The color code is the same as in Fig. \ref{xy6}.}
\label{xpx6}
\end{figure*}

\begin{figure*}[!tH]
\centering
\resizebox{\hsize}{!}{\includegraphics{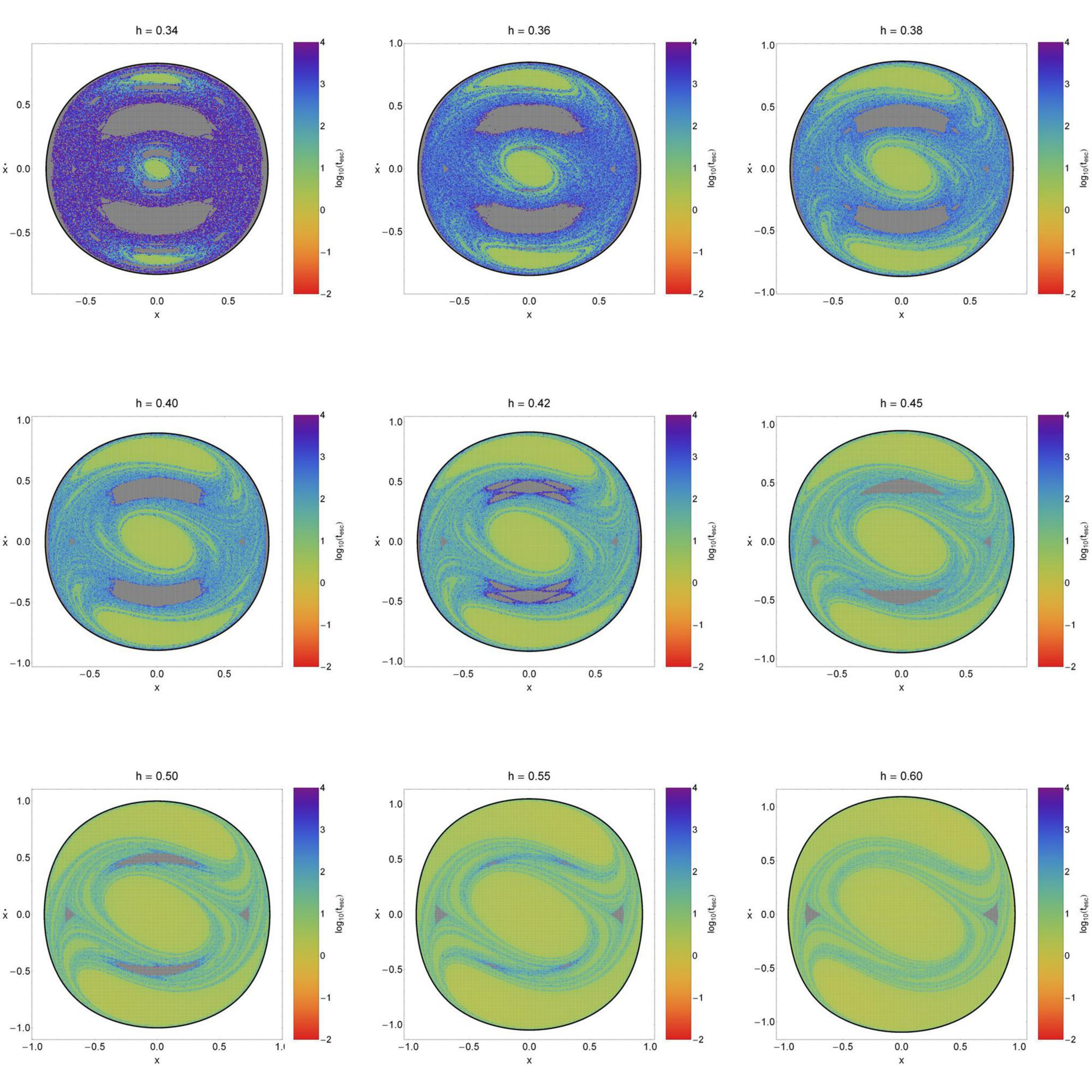}}
\caption{Distribution of the escape times $t_{\rm esc}$ of the orbits on the $(x,\dot{x})$ plane. The darker the color, the larger the escape time. Trapped and non-escaping orbits are indicated by gray color.}
\label{xpxt6}
\end{figure*}

\begin{figure*}[!tH]
\centering
\resizebox{0.8\hsize}{!}{\includegraphics{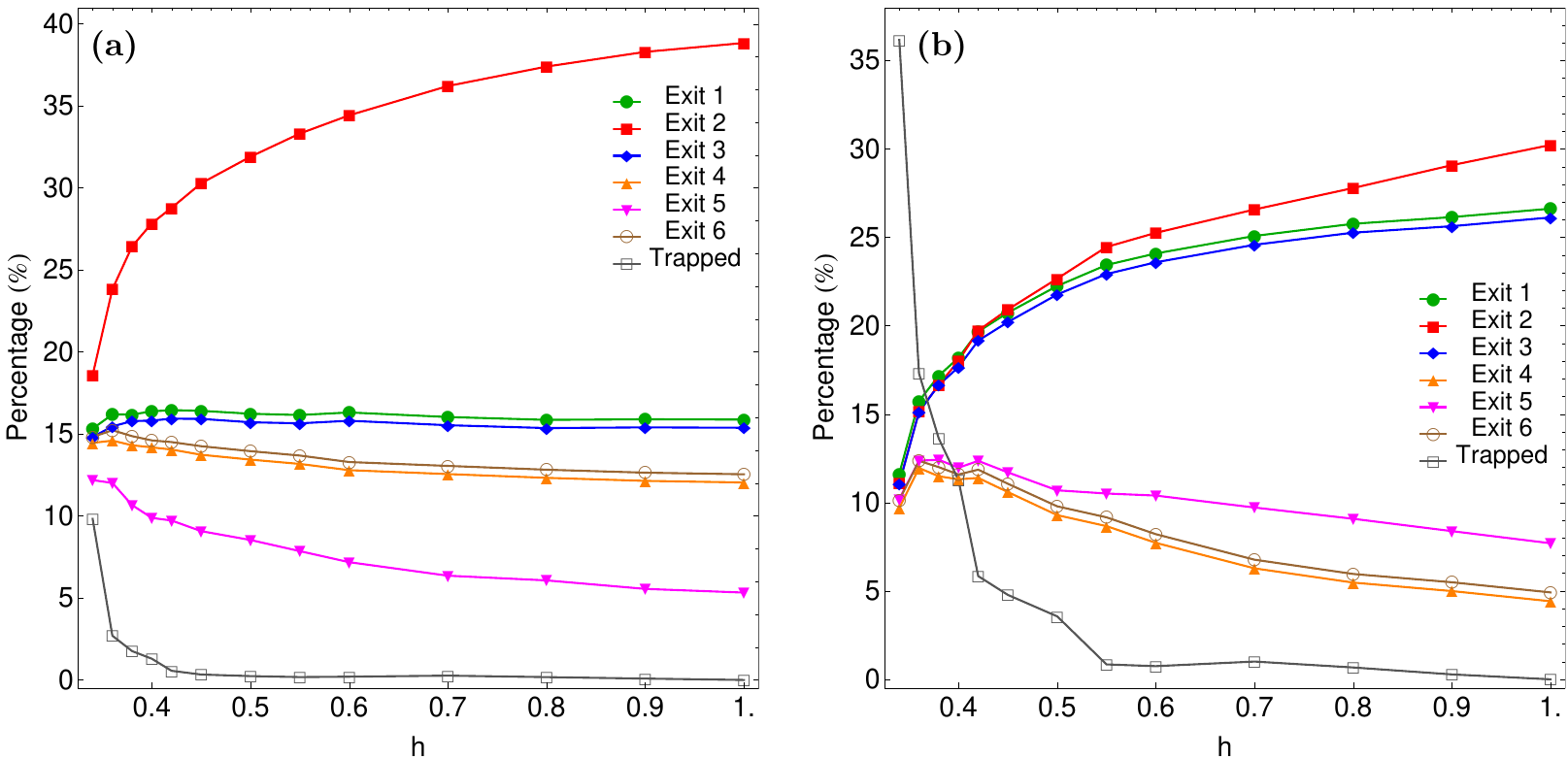}}
\caption{Evolution of the percentages of trapped and escaping orbits when varying the energy $h$ (a-left): on the configuration $(x,y)$ plane and (b-right): on the phase $(x,\dot{x})$ plane.}
\label{percs6}
\end{figure*}

In this case, we shall investigate the escape dynamics of unbounded motion of test particles for values of energy in the set $h$ = \{0.34, 0.36, 0.38, 0.40, 0.42, 0.45, 0.50, 0.55, 0.60\}. We begin with initial conditions of orbits in the configuration $(x,y)$ plane. The orbital structure of the configuration plane for different values of the energy $h$ is show in Fig. \ref{xy6}. Again, following the approach of the previous case, each initial condition is colored according to the escape channel through which the particular orbit escapes. Stability islands on the other hand, filled with initial conditions of ordered orbits which do not escape are indicated as gray regions, while trapped chaotic orbits are shown in black. We observe that things are quite similar to that discussed previously in Fig. \ref{xy5}. In fact, for energy levels very close to the escape energy, the central region of the $(x,y)$ plane is highly fractal and it is also occupied by several stability islands mainly situated at the outer parts of the plane. However, as we increase the value of the energy the regions of regular non-escaping orbits are reduced, the configuration plane becomes less and less fractal and well-defined basins of escape emerge. Additionally, we see that the area on the $(x,y)$ plane occupied by initial conditions of orbits that escape through exit channel 2 grows rapidly with increasing energy and at high energy levels $(h > 0.50)$ it dominates. Once more, as we discussed earlier in Fig. \ref{xy5}, we observe in channel 5 a vertical flow of initial conditions of orbits that escape through exit 2. The following Fig. \ref{xyt6} shows how the escape times $t_{\rm esc}$ of orbits are distributed on the $(x,y)$ plane. Light reddish colors correspond to fast escaping orbits, dark blue/purpe colors indicate large escape periods, while gray color denote trapped orbits. This grid representation of the configuration plane gives us a much more clearer view of the orbital structure and especially about the trapped and non-escaping orbits. In particular, we see that even for the highest energy level studied, that is when $h = 0.60$, two tiny stability islands are still present in the configuration space.

The structure of the $(x,\dot{x})$ phase plane for the same set of values of the energy is shown in Fig. \ref{xpx6}. It is worth noticing that in the phase plane the limiting curve is closed but this does not mean that there is no escape. Remember that we decided to choose such perturbation terms that create the escape channels on the configuration $(x,y)$ plane which is a subspace of the entire four-dimensional $(x,y,\dot{x},\dot{y})$ phase space of the system. We observe a similar behavior to that discussed earlier for the configuration $(x,y)$ plane in Fig. \ref{xy6}. Again, we can distinguish in the phase plane fractal regions where the prediction of the particular escape channel is impossible and regions occupied by escape basins. For low values of the energy $(h < 0.38)$ we can identify initial conditions of trapped chaotic orbits at the boundaries of the two stability islands on the $\dot{x}$ axis. As we proceed to higher energy levels however, the extent of these stability islands is reduced and at relatively high values of the energy $(h > 0.60)$ they completely disappear. Furthermore, it is also seen that the extent of the escape basins of exits 1, 2 and 3 is significantly grows in size with increasing energy. In this case the limiting curves are close and therefore, there is no flow of initial conditions outwards. Fig. \ref{xpxt6} shows the distribution of the escape times $t_{\rm esc}$ of orbits on the $(x,\dot{x})$ plane. It is evident that orbits with initial conditions inside the exit basins escape from the system after short time intervals, or in other words, they possess extremely small escape periods. On the contrary, orbits with initial conditions located in the fractal domains of the phase plane need considerable amount of time in order to find one of the exits and escape.

\begin{figure*}[!tH]
\centering
\resizebox{0.8\hsize}{!}{\includegraphics{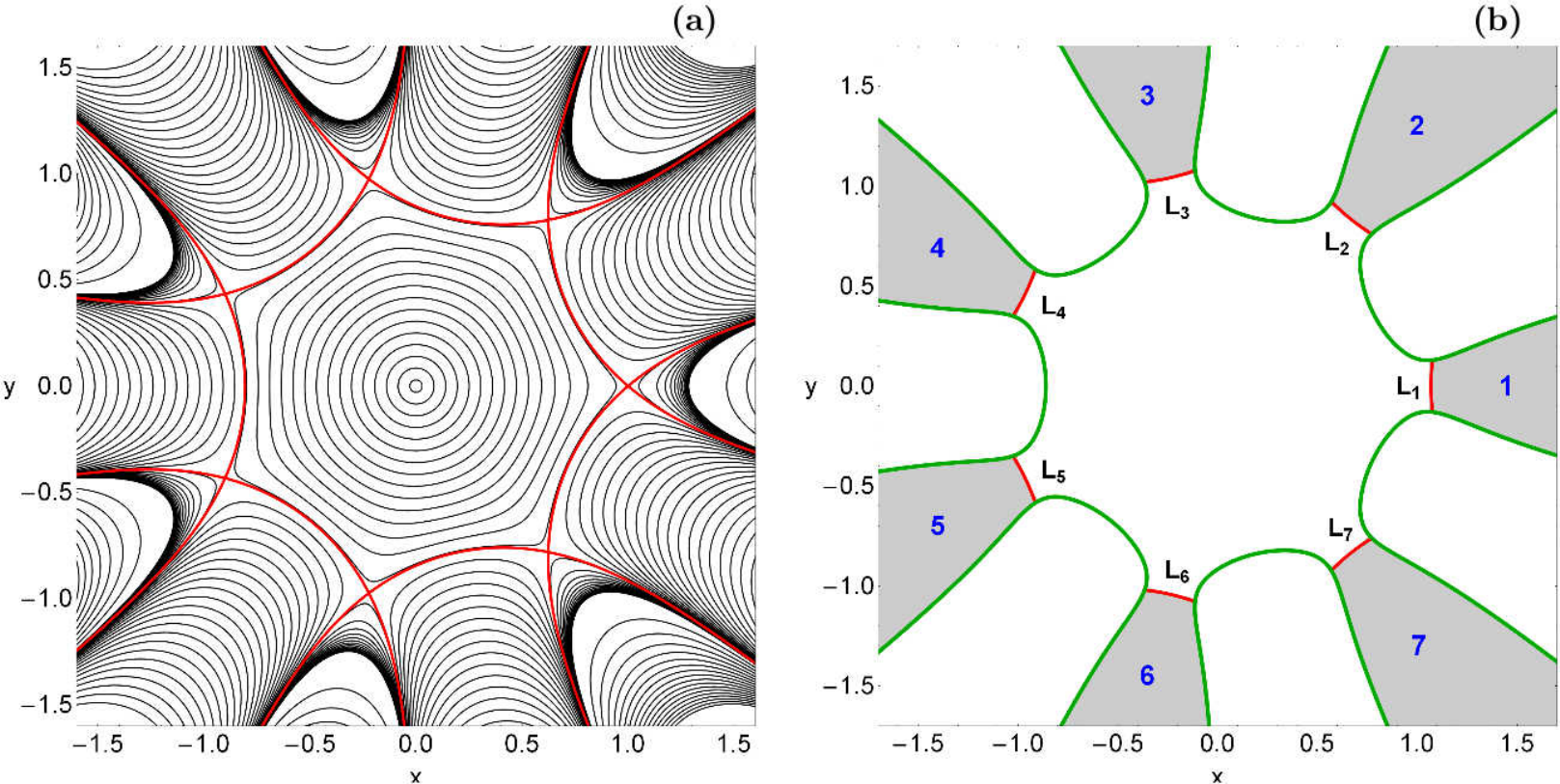}}
\caption{(a-left): Equipotential curves of the total potential (\ref{pot}) for various values of the energy $h$, when seven escape channels are present. The equipotential curve corresponding to the energy of escape is shown with red color; (b-right): The open ZVC at the configuration $(x,y)$ plane when $h = 0.42$. $L_i$, $i=1,...,7$ indicate the seven unstable Lyapunov orbits plotted in red.}
\label{pot7}
\end{figure*}

\begin{figure*}[!tH]
\centering
\resizebox{\hsize}{!}{\includegraphics{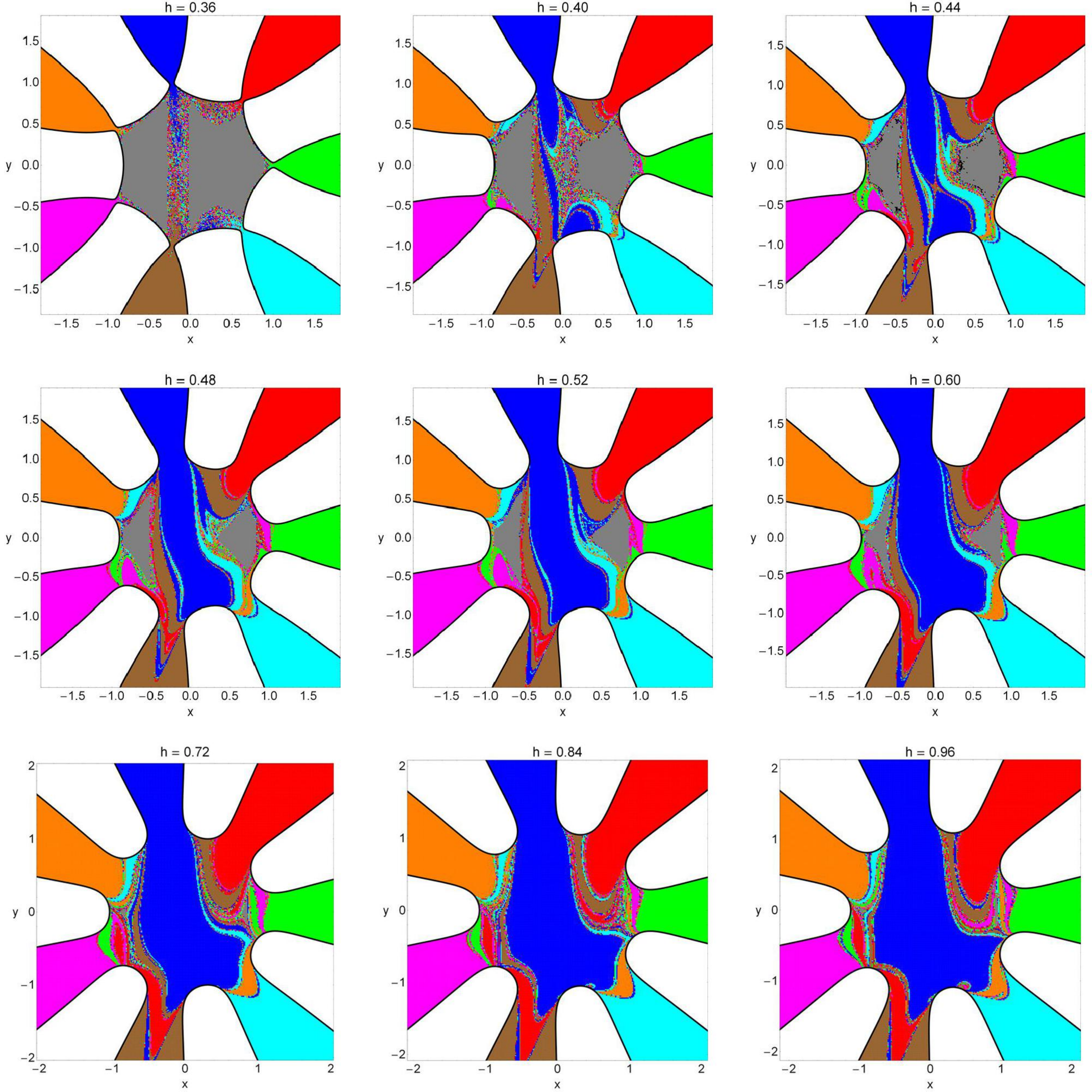}}
\caption{The structure of the configuration $(x,y)$ plane for several values of the energy $h$, distinguishing between different escape channels. The color code is as follows: Escape through channel 1 (green); escape through channel 2 (red); escape through channel 3 (blue); escape through channel 4 (orange); escape through channel 5 (magenta); escape through channel 6 (brown); escape through channel 7 (cyan); non-escaping regular (gray); trapped chaotic (black).}
\label{xy7}
\end{figure*}

The evolution of the percentages of trapped and escaping orbits on the configuration $(x,y)$ plane when the value of the energy $h$ varies is presented in Fig. \ref{percs6}a. It is seen that for $h = 0.34$, that is the first investigated energy level above the escape energy, escaping orbits through channels 1, 3, 4, and 6 share the same percentage (around 15\%), escapers through channel 2 have a slightly elevated percentage (around 18\%), while trapped orbits possess a low rate corresponding only to about 10\% of the configuration plane. Once more, as we increase the value of the energy the rate of trapped orbits decreases and eventually vanishes for $h > 0.8$. Furthermore, we observe that the percentage of escaping orbits through channel 2 grows with increasing energy and remains always the most populated escape channel. The percentages of escaping orbits through channels 1, 3, 4 and 6 on the other hand, are almost unperturbed by the shifting on the orbital energy and the seem to saturate around 15\%, while the rate of escaping orbits through exit 5 displays a gradual decrease. In general terms, we may conclude that throughout the energy range studied, the majority of orbits in the configuration $(x,y)$ plane choose to escape through exit channel 2, while exit 5 seems to be the least favorable among the escape channels. It is evident from Fig. \ref{percs6}b where the evolution of the percentages of trapped and escaping orbits on the phase plane as a function of the value of the energy $h$ is presented that the pattern has many differences comparing to that discussed previously in Fig. \ref{percs6}a. To begin with, we observe that for $h = 0.34$ more than 35\% of the phase plane corresponds to initial conditions of orbits that do not escape, while all the escape channels are equiprobable taking into account that all channels have the same rate thus sharing about 60\% of the phase space. As the value of the energy increases and we move away from the escape energy it is seen that the rate of trapped orbits is heavily reduced, while the percentages of the escape channels start to diverge following two different patterns. Being more specific, one may observe that the rates of escaping orbits through exits 1, 2 and 3 start to grow, while on the other hand the percentages of escapers through exits 4, 5 and 6 exhibit a gradual decrease. At the highest energy level studied $(h = 1.0)$, about 30\% of the total orbits escape through channel 2, exit channels 1 and 3 share about half of the phase plane, while exit channels 4 and 6 share about 10\% of the same plane. Thus, one may reasonably conclude that throughout the energy range studied, the vast majority of orbits in the phase $(x,\dot{x})$ plane choose to escape through channels 1, 2 and 3, while channels 4, 5 and 6 are much less likely to be chosen.

\subsection{Case III: Seven channels of escape}
\label{case3}

Our escape quest continues considering a Hamiltonian system with seven exit channels where the escape energy is equal to 5/14. In order to obtain this number of exits $(n = 7)$ in the limiting curve in the configuration $(x,y)$ plane, the perturbation term should be
\begin{equation}
V_1(x,y) = - \frac{1}{7}\left(x^7 - 21 x^5 y^2 + 35 x^3 y^4 - 7 x y^6 \right),
\label{ham3}
\end{equation}
according to the first generating function of Eqs. (\ref{gens}). We observe that the corresponding Hamiltonian $H = H_0 + H_1$ is symmetric with respect to $y \rightarrow - y$. In Fig. \ref{pot7}a we see the equipotential curves of the total potential (\ref{pot}) for various values of the energy $h$, while the equipotential corresponding to the energy of escape $h_{esc}$ is plotted with red color in the same plot. Furthermore, the open ZVC at the configuration $(x,y)$ plane when $h = 0.42 > h_{esc}$ is presented with green color in Fig. \ref{pot6}b and the seven channels of escape are shown. In the same figure, the seven unstable Lyapunov orbits $L_i$, $i = {1,...,7}$ are denoted using red color.

\begin{figure*}[!tH]
\centering
\resizebox{\hsize}{!}{\includegraphics{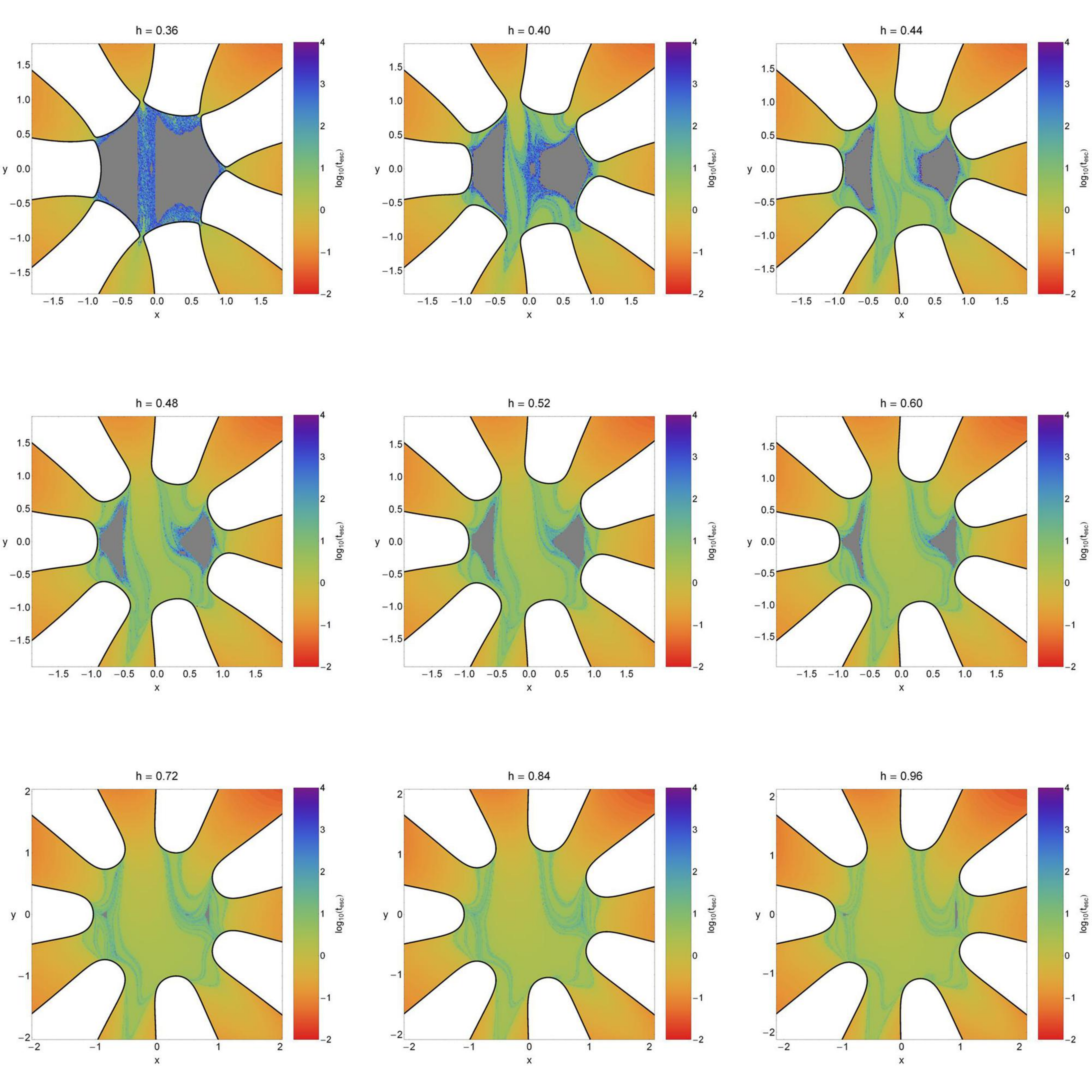}}
\caption{Distribution of the escape times $t_{\rm esc}$ of the orbits on the $(x,y)$ plane. The darker the color, the larger the escape time. Trapped and non-escaping orbits are indicated by gray color.}
\label{xyt7}
\end{figure*}

\begin{figure*}[!tH]
\centering
\resizebox{\hsize}{!}{\includegraphics{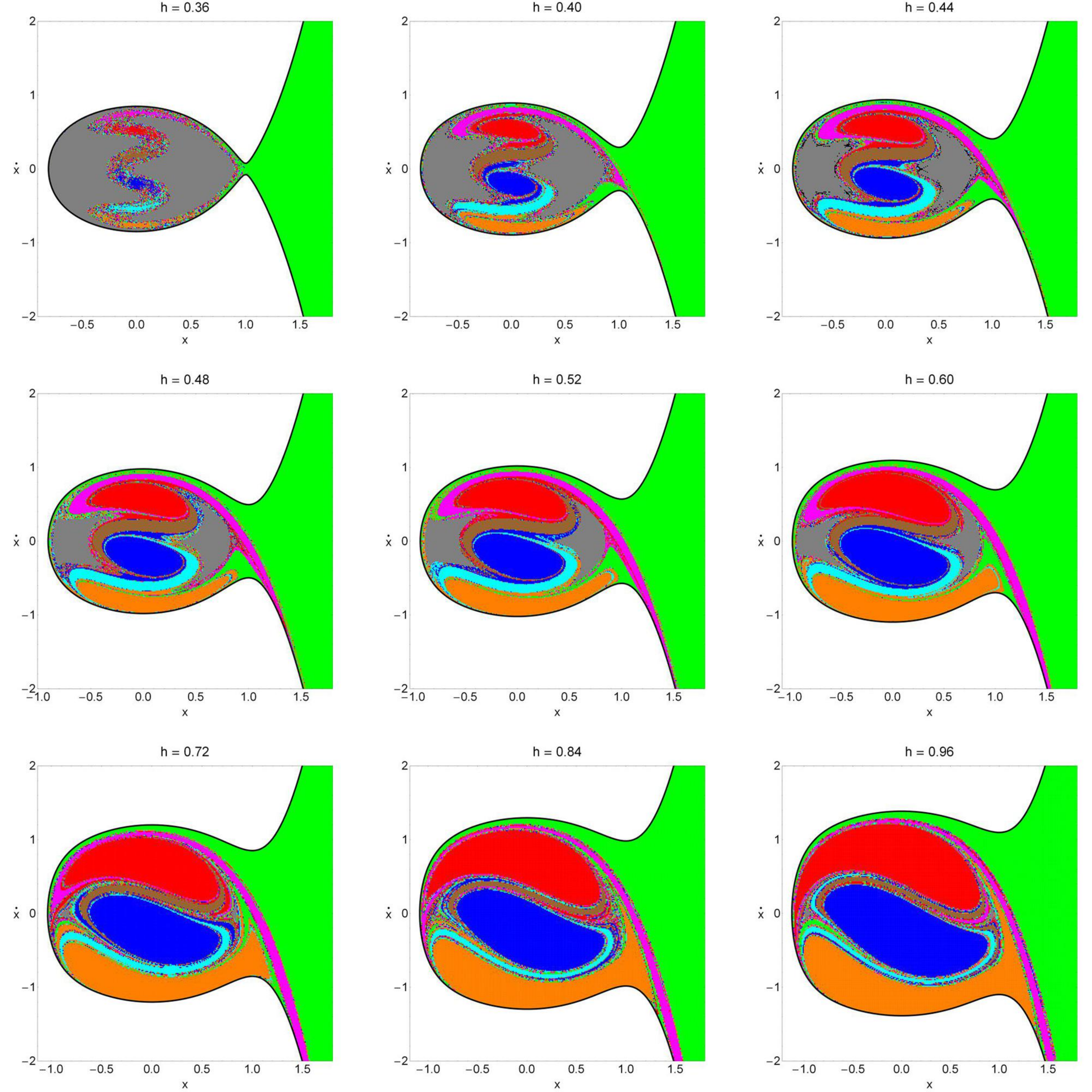}}
\caption{The structure of the phase $(x,\dot{x})$ plane for several values of the energy $h$, distinguishing between different escape channels. The color code is the same as in Fig. \ref{xy7}.}
\label{xpx7}
\end{figure*}

\begin{figure*}[!tH]
\centering
\resizebox{\hsize}{!}{\includegraphics{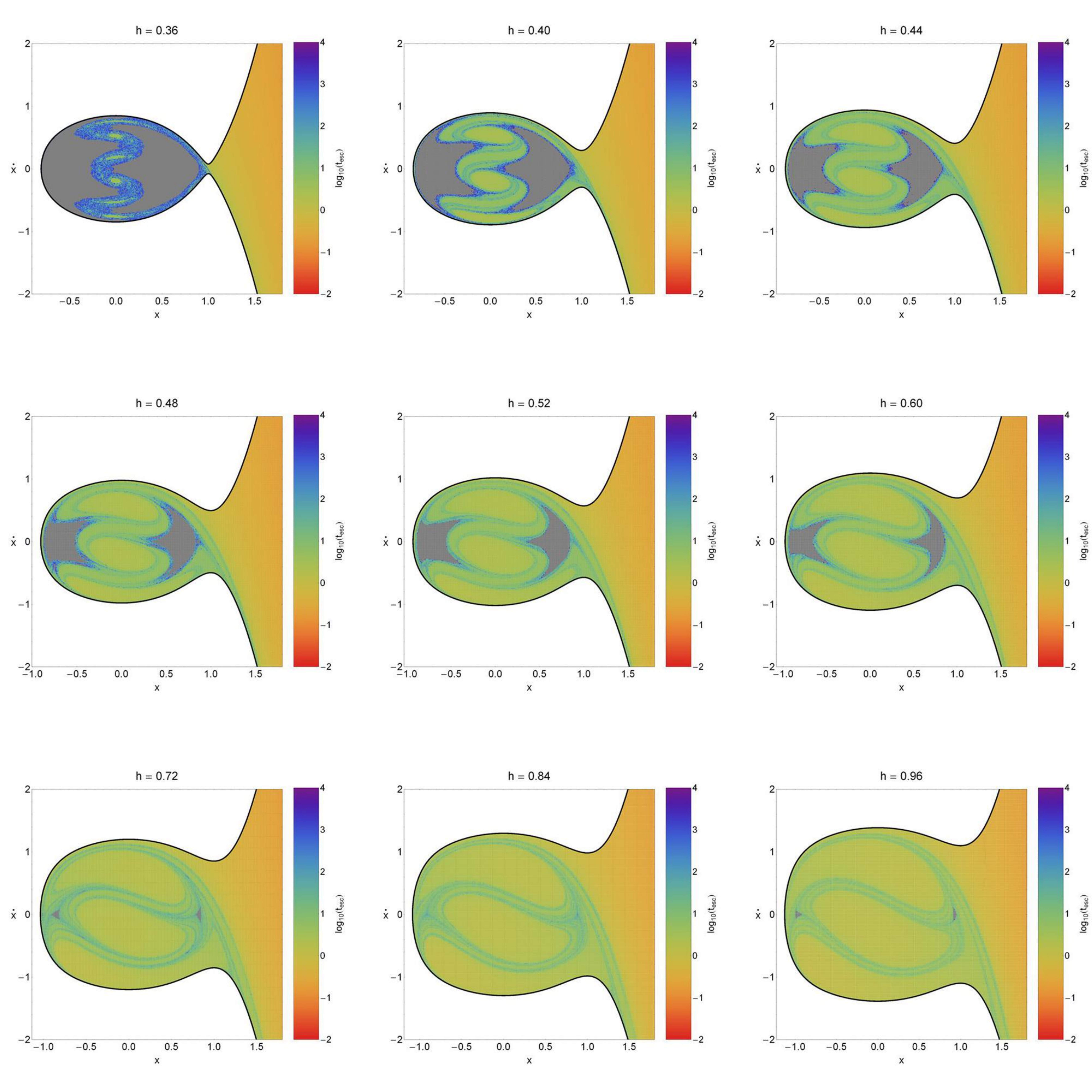}}
\caption{Distribution of the escape times $t_{\rm esc}$ of the orbits on the $(x,\dot{x})$ plane. The darker the color, the larger the escape time. Trapped and non-escaping orbits are indicated by gray color.}
\label{xpxt7}
\end{figure*}

\begin{figure*}[!tH]
\centering
\resizebox{0.8\hsize}{!}{\includegraphics{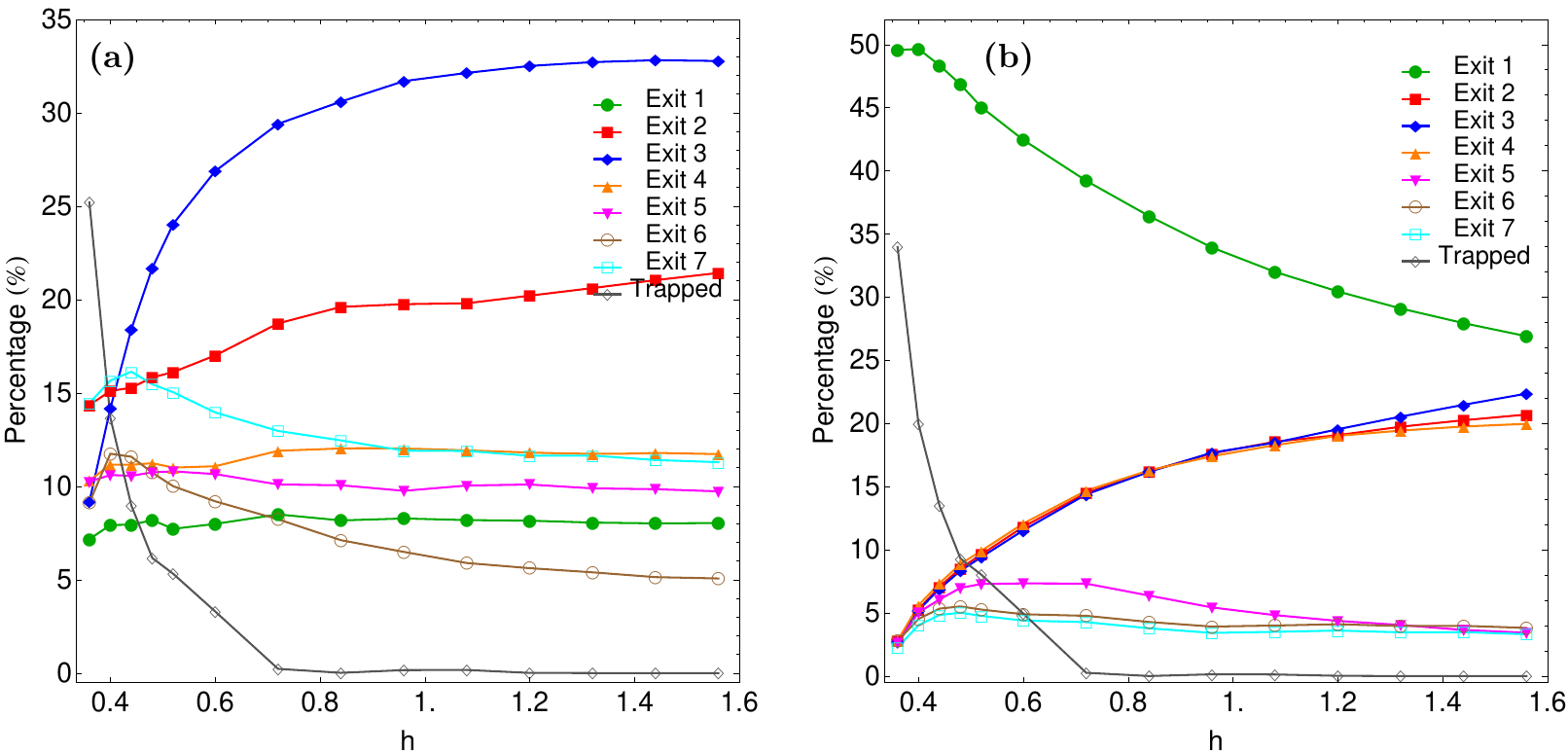}}
\caption{Evolution of the percentages of trapped and escaping orbits when varying the energy $h$ (a-left): on the configuration $(x,y)$ plane and (b-right): on the phase $(x,\dot{x})$ plane.}
\label{percs7}
\end{figure*}

\begin{figure*}[!tH]
\centering
\resizebox{0.8\hsize}{!}{\includegraphics{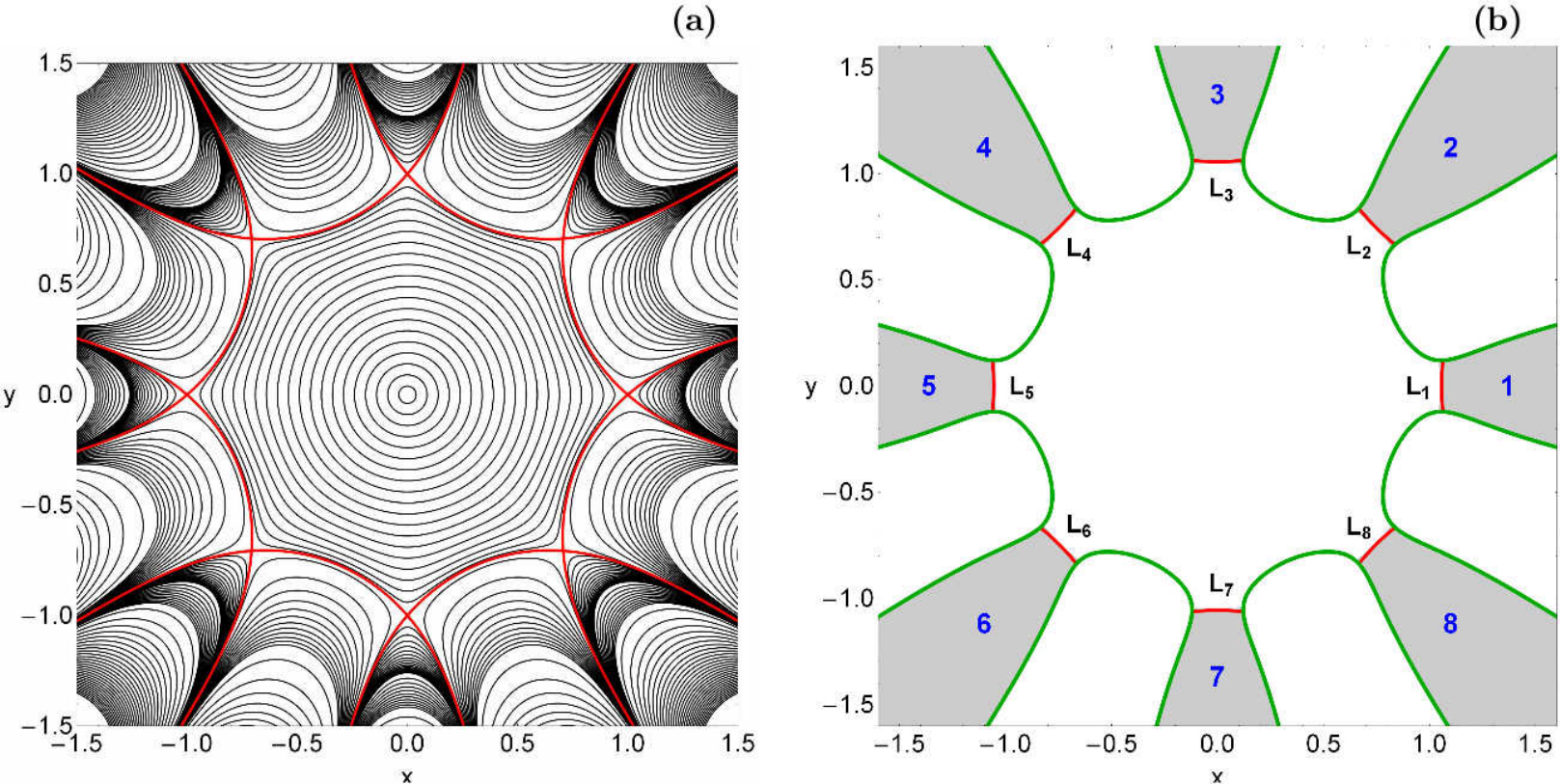}}
\caption{(a-left): Equipotential curves of the total potential (\ref{pot}) for various values of the energy $h$, when eight escape channels are present. The equipotential curve corresponding to the energy of escape is shown with red color; (b-right): The open ZVC at the configuration $(x,y)$ plane when $h = 0.44$. $L_i$, $i=1,...,8$ indicate the eight unstable Lyapunov orbits plotted in red.}
\label{pot8}
\end{figure*}

\begin{figure*}[!tH]
\centering
\resizebox{\hsize}{!}{\includegraphics{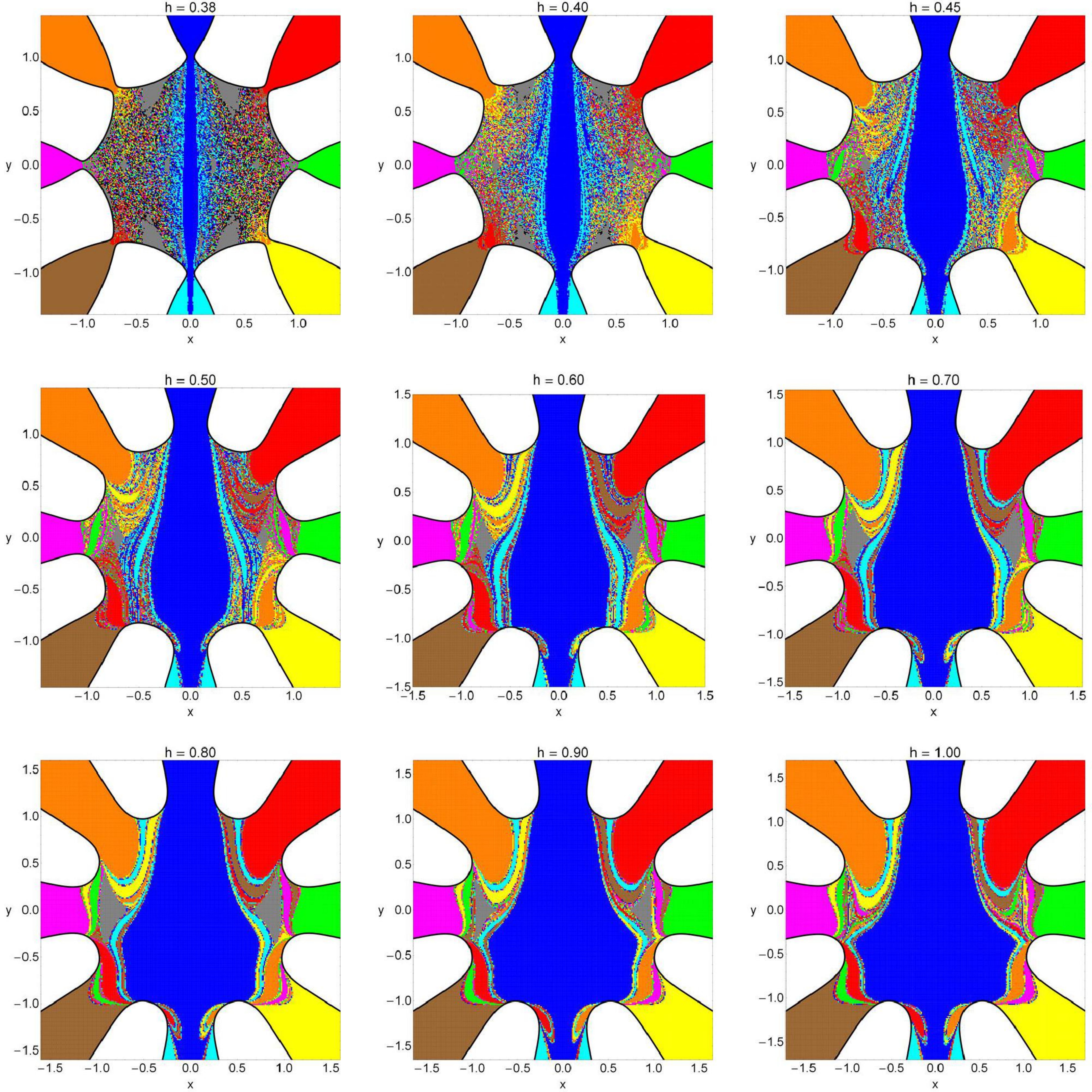}}
\caption{The structure of the configuration $(x,y)$ plane for several values of the energy $h$, distinguishing between different escape channels. The color code is as follows: Escape through channel 1 (green); escape through channel 2 (red); escape through channel 3 (blue); escape through channel 4 (orange); escape through channel 5 (magenta); escape through channel 6 (brown); escape through channel 7 (cyan); escape through channel 8 (yellow); non-escaping regular (gray); trapped chaotic (black).}
\label{xy8}
\end{figure*}

In this case, the set of values of the total orbital energy of the test particles is $h$ = \{0.36, 0.40, 0.44, 0.48, 0.52, 0.60, 0.72, 0.84, 0.96\}. First, we consider initial conditions of orbits in the configuration $(x,y)$ plane and in Fig. \ref{xy7} the orbital structure of the configuration plane for different values of the energy $h$ is presented. As in all previous cases, each initial condition is colored according to the escape channel through which the test particle escapes. For $h = 0.36$, that is an energy level just above the critical escape energy, the vast majority of the interior region of the configuration plane is covered by initial conditions of non-escaping regular orbits forming two stability islands which are separated by a highly fractal layer. As we increase the value of the energy the stability islands are reduced, while at the same time the fractality of the configuration plane is considerably reduced and well-formed basis of escape emerge. Furthermore, we see that the area on the $(x,y)$ plane occupied by initial conditions of orbits that escape through exit channel 3 grows rapidly with increasing energy and at high energy levels $(h > 0.60)$ they dominate. The outwards flow of initial conditions is once more present in channel 6. The distribution of the escape times $t_{\rm esc}$ of orbits on the configuration plane is given in Fig. \ref{xyt7}, where light reddish colors correspond to fast escaping orbits, dark blue/purpe colors indicate large escape periods, while gray color denote trapped and non-escaping orbits. It is evident that the two main stability islands even though they reduce in size with increasing energy they do not completely disappear since for $h = 0.96$  we still observe the presence of two tine stability islands inside the interior region of the configuration plane.

The following Fig. \ref{xpx7} shows the orbital structure of the $(x,\dot{x})$ phase plane for the same set of values of the total energy $h$. It is seen that the phase space is divided into three types of regions: (i) regions of regular motion where the corresponding orbits do not escape; (ii) fractal regions where we cannot predict the particular escape channel for a given orbit and (iii) regions where the initial conditions of orbits define broad basins of escape. The first and the second type of regions occupy large portion of the phase space for low values of the energy $(h < 0.50)$, while for larger values their extent is considerably confined. The third type on the other hand, exhibits the complete opposite behavior. In particular, we observe that the basins of escape corresponding to exits 2, 3 and 4 grow significantly in size with increasing energy. Similarly as in Fig. \ref{xpx5}, a weak stream of initial conditions of orbits is identified in the phase planes. Fig. \ref{xpxt7} depicts the distribution of the escape times $t_{\rm esc}$ of orbits on the phase $(x,\dot{x})$ plane. Once more we see that orbits with initial conditions inside the basins of escape have very small escape periods and therefore, they escape to infinity quite early. On the contrary, orbits with situated in the fractal regions of the phase plane require long time intervals in order to find one of the exits and escape.

\begin{figure*}[!tH]
\centering
\resizebox{\hsize}{!}{\includegraphics{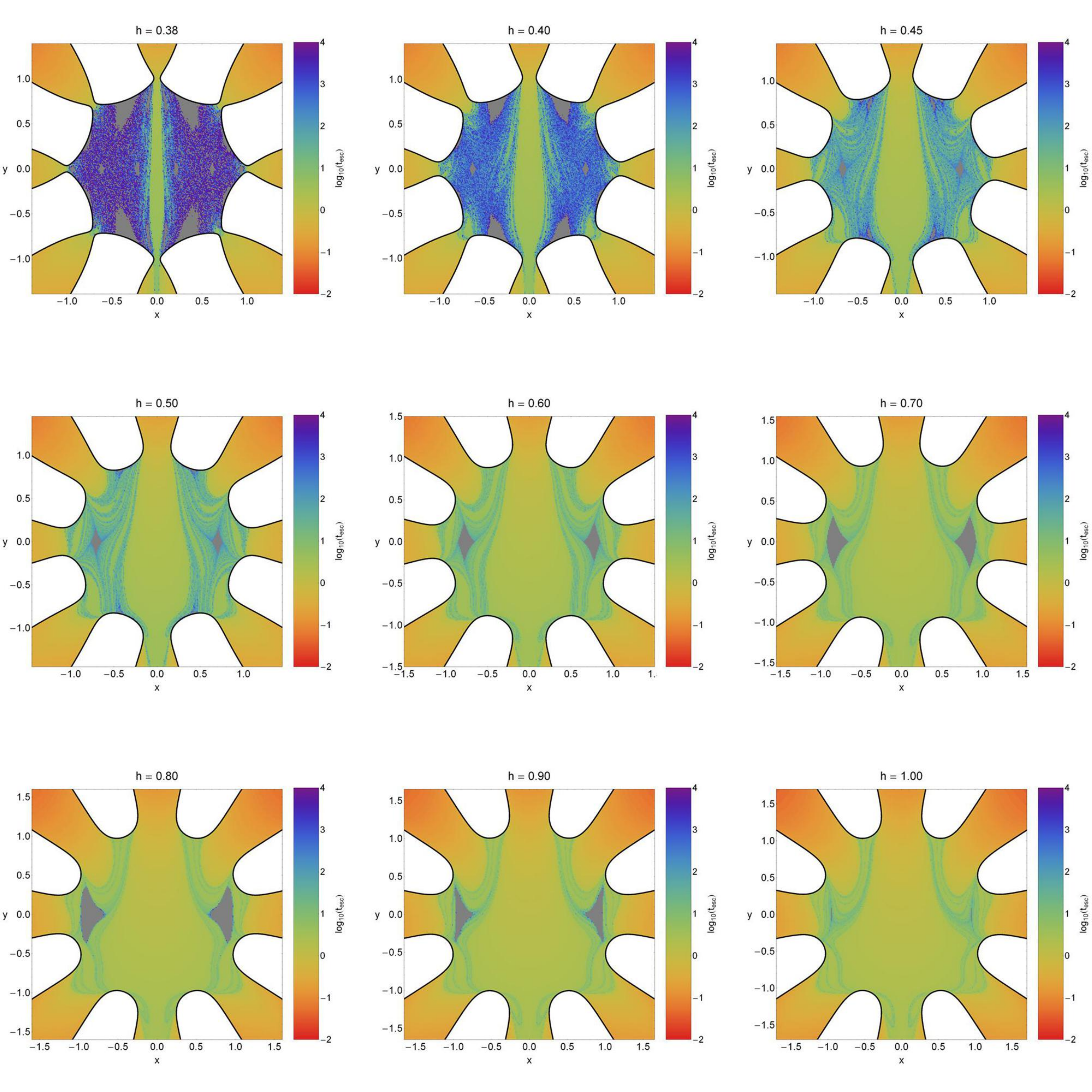}}
\caption{Distribution of the escape times $t_{\rm esc}$ of the orbits on the $(x,y)$ plane. The darker the color, the larger the escape time. Trapped and non-escaping orbits are indicated by gray color.}
\label{xyt8}
\end{figure*}

\begin{figure*}[!tH]
\centering
\resizebox{\hsize}{!}{\includegraphics{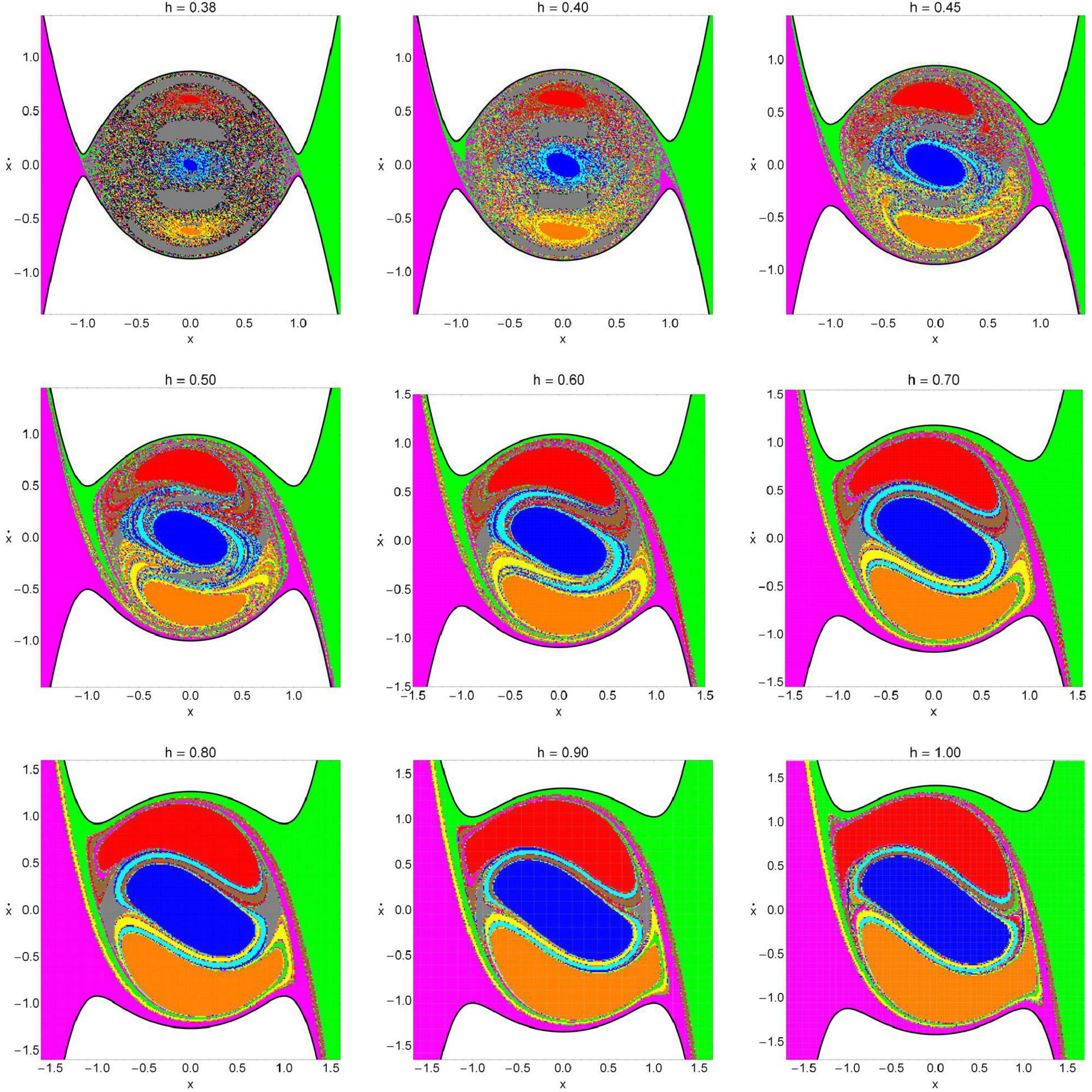}}
\caption{The structure of the phase $(x,\dot{x})$ plane for several values of the energy $h$, distinguishing between different escape channels. The color code is the same as in Fig. \ref{xy8}.}
\label{xpx8}
\end{figure*}

\begin{figure*}[!tH]
\centering
\resizebox{\hsize}{!}{\includegraphics{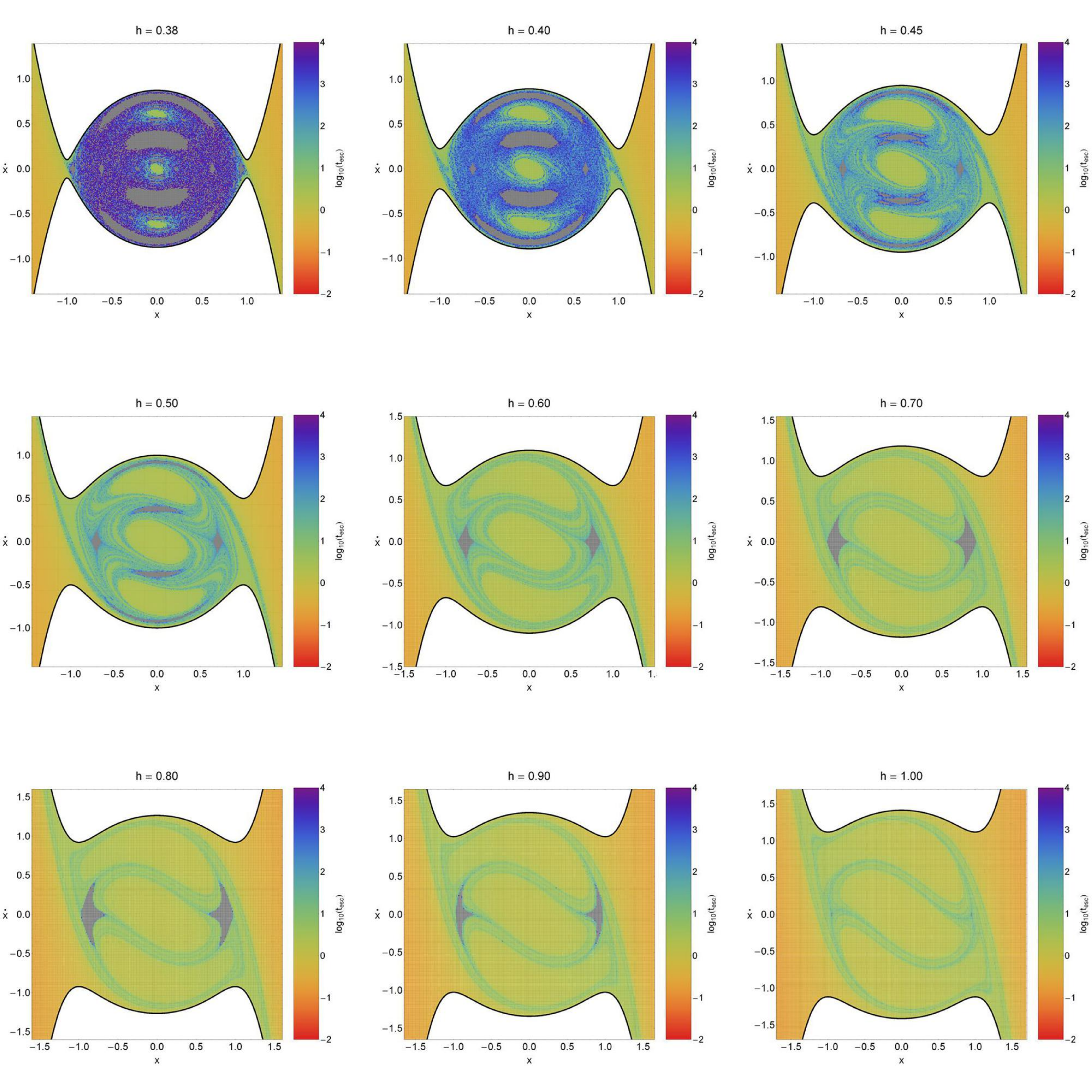}}
\caption{Distribution of the escape times $t_{\rm esc}$ of the orbits on the $(x,\dot{x})$ plane. The darker the color, the larger the escape time. Trapped and non-escaping orbits are indicated by gray color.}
\label{xpxt8}
\end{figure*}

\begin{figure*}[!tH]
\centering
\resizebox{0.8\hsize}{!}{\includegraphics{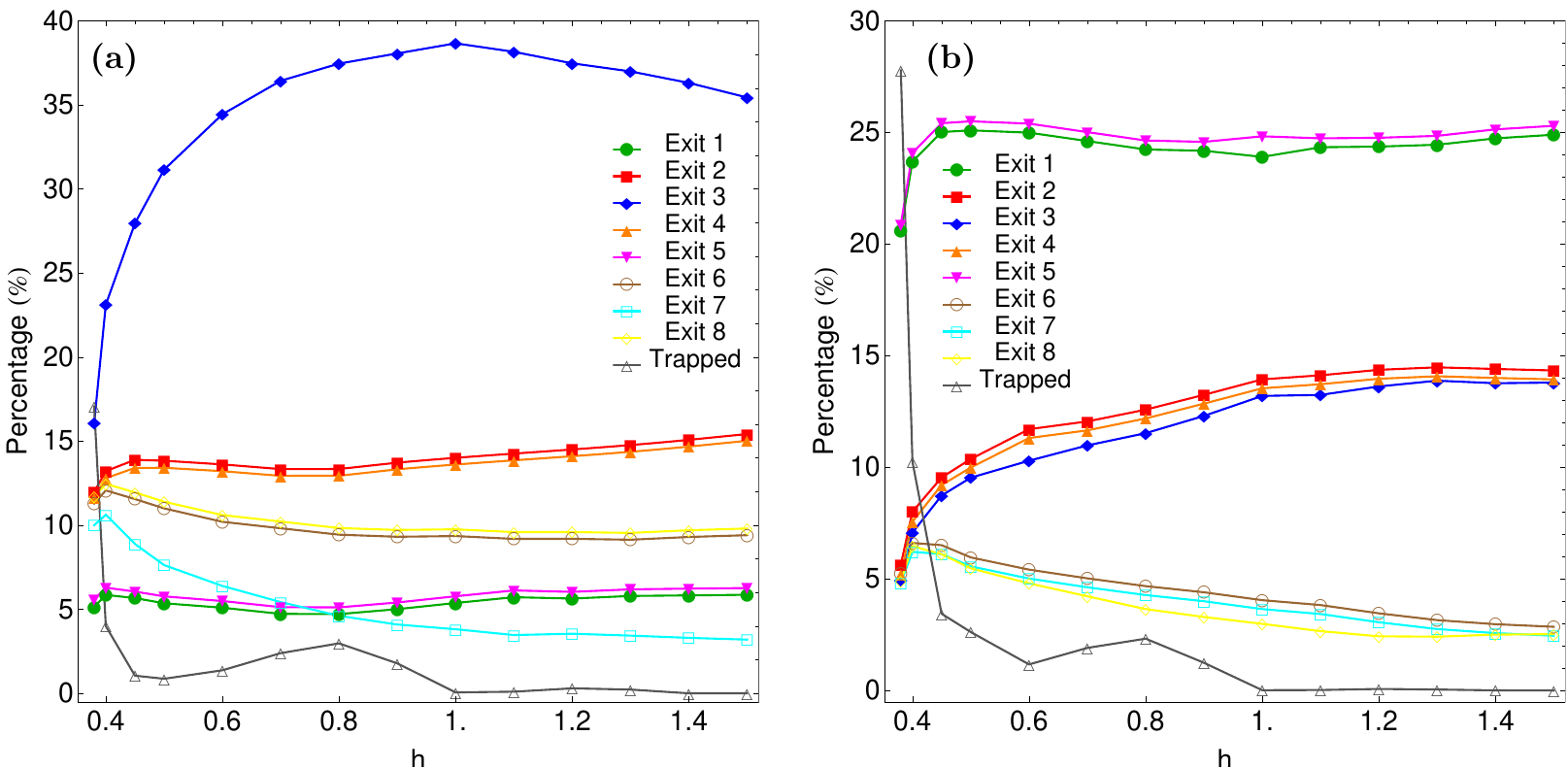}}
\caption{Evolution of the percentages of trapped and escaping orbits when varying the energy $h$ (a-left): on the configuration $(x,y)$ plane and (b-right): on the phase $(x,\dot{x})$ plane.}
\label{percs8}
\end{figure*}

In Fig. \ref{percs7}a we see the evolution of the percentages of trapped and escaping orbits on the configuration $(x,y)$ plane when the value of the energy $h$ varies. For $h = 0.36$, that is an energy level just above the escape energy $h_{esc}$, trapped orbits occupy about one fourth of the entire configuration plane. In addition, the percentages of escaping orbits through exits 2 and 7 share about 30\% of the $(x,y)$ plane, while all the other rates apart form that of exit 1 have about the same value around 10\%. The portion of trapped orbits reduces as the value of the energy increases and for $h > 1$ they vanish. The percentage of escaping orbits through exit channel 3 on the other hand increases as we proceed to higher energy levels and for $h > 0.4$ it is the most populated type of orbits. The rate of escaping orbits through exit 2 also increases and for high values of energy it seems to saturate around 22\%, while that of exit 6 decreases reaching 5\% at the highest energy level studied. The percentages of all the remaining exit channels seem to be almost unperturbed by the change on the value of the energy holding values around 10\% throughout. Therefore, one may conclude that in the configuration $(x,y)$ plane the majority of orbits choose to escape either through exit channel 2, or exit 3, while all the other exits are significantly less probable to be chosen by the test particles. In the same vein, we present in Fig. \ref{percs7}b the evolution of the percentages of trapped and escaping orbits on the phase plane as a function of the value of the energy $h$. Here it is evident that things are quite different. At the lowest examined energy level $(h = 0.36)$ it is found that half of the phase space is occupied by initial conditions of orbits that escape through channel 1, about 35\% of the integrated initial conditions correspond to trapped regular orbits, while the remaining 15\% of the phase plane is shared by escaping orbits through channels 2 to 7. As the value of the energy increases the percentages of both trapped and escaping through exit 1 orbits are reduced however, the latter type of orbits remains throughout the most populated one. At the same time, the percentages of escaping orbits through channels 2 to 7 start to diverge and produce two distinct branches. The first branch contains the evolution of the rates of escaping orbits through exits 2, 3 and 4 which all of them exhibit a common increase and at the highest energy level studied $(h = 1.56)$ they share about 60\% of the phase space. The second branch includes the percentages of escaping orbits through channels 5, 6 and 7 and we see that all of them are almost unperturbed by the energy shifting evolving at low values less than 5\%. Taking into account the above-mentioned results regarding the phase $(x,\dot{x})$ space we may say that throughout the energy range studied, the vast majority of orbits choose to escape through one of the first four channels (exits 2, 3 and 3 are practically equiprobable), while the remaining channels (5 to 7) are significantly less likely to be chosen.

\subsection{Case IV: Eight channels of escape}
\label{case4}

The last case under investigation is the case where the Hamiltonian system has eight channels of escape $(n = 8)$. The corresponding perturbation function is obtained from the second generating function (\ref{gens}) and it reads
\begin{equation}
V_1(x,y) = - \frac{1}{8}\left(x^8 - 28 x^6 y^2 + 70 x^4 y^4 - 28 x^2 y^6 + y^8 \right),
\label{ham1}
\end{equation}
while the corresponding escape energy is equal to 3/8. The total Hamiltonian $H = H_0 + H_1$ is invariant under $x \rightarrow - x$ and/or $y \rightarrow - y$. The equipotential curves of the total potential (\ref{pot}) for various values of the energy $h$ are shown in Fig. \ref{pot8}a. The equipotential corresponding to the energy of escape is plotted with red color in the same plot. The open ZVC at the configuration $(x,y)$ plane when $h = 0.44 > h_{esc}$ is presented with green color in Fig. \ref{pot5}b and the eight channels of escape are shown. In the same plot, we denote the eight unstable Lyapunov orbits by $L_i$, $i = 1,...,8$ using red color.

The escape properties and mechanism of unbounded motion of test particles for values of energy in the set $h$ = \{0.38, 0.40, 0.45, 0.50, 0.60, 0.70, 0.80, 0.90, 1.00\} will be examined. We begin, as usual, with initial conditions of orbits in the configuration $(x,y)$ plane. Fig. \ref{xy8} shows the orbital structure of the configuration plane for different values of the energy $h$. Again, following the same approach as in all previous cases, each initial condition is colored according to the escape channel through which the particular orbit escapes. Areas corresponding to non-escaping regular orbits on the other hand, are indicated as gray regions, while trapped chaotic orbits are shown in black. We see that for values of energy very close to the escape energy $(h < 0.50)$ the majority of the central region of the configuration plane is fractal, some small stability islands are present situated mainly at the outer parts of the interior region, while we observe a week stream of initial conditions of orbits that escape from channel 3 which crosses vertically the $(x,y)$ plane and flows outwards from channel
7 (a similar phenomenon was also observed in all previous cases). As the value of the energy increases this stream becomes more and more strong evolving to a wide basin of escape which eventually takes over most of the interior region of the configuration plane. Moreover, additional smaller basins of escape emerge mainly around the unstable Lyapunov orbits, while the stability islands containing the initial conditions of non-escaping regular orbits are reduced in size. Here we should like to note that in general terms, throughout the energy range the structure of the configuration $(x,y)$ plane is somehow symmetrical with respect to the $x = 0$ axis. The distribution of the escape times $t_{\rm esc}$ of orbits on the configuration plane is given in Fig. \ref{xyt8}. Light reddish colors correspond to fast escaping orbits, dark blue/purpe colors indicate large escape periods, while gray color denote trapped and non-escaping orbits orbits. Here, we have a better view regarding the amount of trapped orbits. We see that for $h = 1.0$ all orbits escape from the system. Moreover we observe that orbits with initial conditions close to the area occupied by trapped orbits have significantly high escape periods, while on the other hand, orbits located near the exit channels escape very quickly having escaping rates of about two orders smaller.

We proceed with the phase $(x,\dot{x})$ plane, the structure of which for the same set of values of the energy is presented in Fig. \ref{xpx8}. It is seen that this time the limiting curve is open at both sides. One may observe that for $h < 0.4$ a large part of the phase plane is covered by several sets of stability islands corresponding to non-escaping orbits, the remaining has a highly fractal structure, while only three basins of escape are shown; one in the central region of the phase plane and two other one above and one below it. However, as the value of the energy increases and we move far away for the escape energy, the extent of these three basins of escape grows and for $h > 0.70$ they dominate. At the same time, small elongated spiral basins of escape emerge inside the fractal region which surrounds the central escape basin. Furthermore, at very high energy levels $(h > 1.0)$ we see that non-escaping regular orbits disappear completely from the grid and the three main basins of escape take over the vast majority of the phase plane, while the elongated escape basins remain confined to the central region. As we noticed previously when discussing the configuration $(x,y)$ plane, there is also a symmetry in the phase plane. In particular, throughout the energy range the structure of the phase plane $(x,\dot{x})$ is somehow symmetrical (though not with the strick sense) with respect to the $\dot{x} = 0$ axis. In this case, the limiting curves in the phase plane are open in both sides thus, we observe the existence of two streams of initial conditions leaking out and extend to infinity. The following Fig. \ref{xpxt8} shows the distribution of the escape times $t_{\rm esc}$ of orbits on the phase $(x,\dot{x})$ plane. It is clear that orbits with initial conditions inside the exit basins escape to infinity after short time intervals, or in other words, they possess extremely small escape periods. On the contrary, orbits with initial conditions located in the fractal parts of the phase plane need considerable amount of time in order to find one of the four exits and escape. It is seen that at the highest energy level studied $(h = 1.0)$ there is no indication of bounded motion and all orbits escape to infinity sooner or later.

It is of particular interest to monitor the evolution of the percentages of trapped and escaping orbits on the configuration $(x,y)$ plane when the value of the energy $h$ varies. A diagram depicting this evolution is presented in Fig. \ref{percs8}a. We see that for $h = 0.38$, that is an energy level just above the escape energy, about 17\% of the configuration plane is covered by initial conditions of trapped orbits. As the value of the energy increases however, the rate of trapped orbits drops rapidly and eventually at $h > 1.0$ it vanishes. We also observe that the evolution of the percentages of orbits escaping through channels 1, 2 and 6 coincide with the evolution of the percentages escaping through channels 5, 4 and 8, respectively. We anticipated this behaviour of the escape percentages, which is a natural result of the symmetrical structure of the configuration $(x,y)$ plane. We also anticipated the domination of escaping orbits through exit 3 due to the strong escape stream. It is seen that initially $(h = 0.38)$ the rates of escaping orbits through exits 2, 4, 6, 7 and 8 coincide at about 12\%. Then, with increasing energy the rates of these types of escaping orbits start to diverge but only escapers through exit 7 decrease; all the others remain almost unperturbed. At the highest energy studied, escaping orbits through channels 2 and 4 share about 30\% of the configuration plane, escaping orbits through channels 6 and 8 share about 20\% of the same plane, while escaping orbits trough channels 1 and 5 occupy only about 12\% of the grid. Therefore, one may reasonably conclude that in general terms, throughout the range of the values of the energy studied, the majority of orbits in the configuration $(x,y)$ plane choose to escape through channels 2, 3 and 4. The evolution of the percentages of trapped and escaping orbits on the phase plane as a function of the value of the energy $h$ is given in Fig. \ref{percs8}b. For $h = 0.38$, we see that trapped orbits is the most populated type of orbits as they occupy about 27\% of the phase space. However as usual, with increasing energy the dominance of trapped orbits deteriorates rapidly due to the increase of the rates of escaping orbits forming basins of escape. We observe that the evolution of several rates of escaping orbits coincide due to the symmetry of the phase space. In particular, there are three different branches, given with decreasing strength (rate): (i) containing the rates of exits 1 and 5; (ii) containing the rates of exits 2, 3 and 4; (iii) containing the rates of exits 6, 7 and 8. The first branch seems to be unaffected by the change on the value of the energy and exits 1 and 5 share about half of the phase space throughout. The rates of escaping orbits that belong to the second branch exhibit a small increase and an the highest energy level studied they share about 40\% of the grid. The percentages of escaping orbits that belong to the third branch on the other hand, are reduced and when $h = 1.5$ they share the remaining 10\% of the phase plane. Thus, we may conclude that the vast majority of orbits in the phase $(x,\dot{x})$ plane displays clear sings of preference through exits 1 and 5, while channels 6, 7 and 8 have considerable less probability to be chosen.

\begin{figure*}[!tH]
\centering
\resizebox{\hsize}{!}{\includegraphics{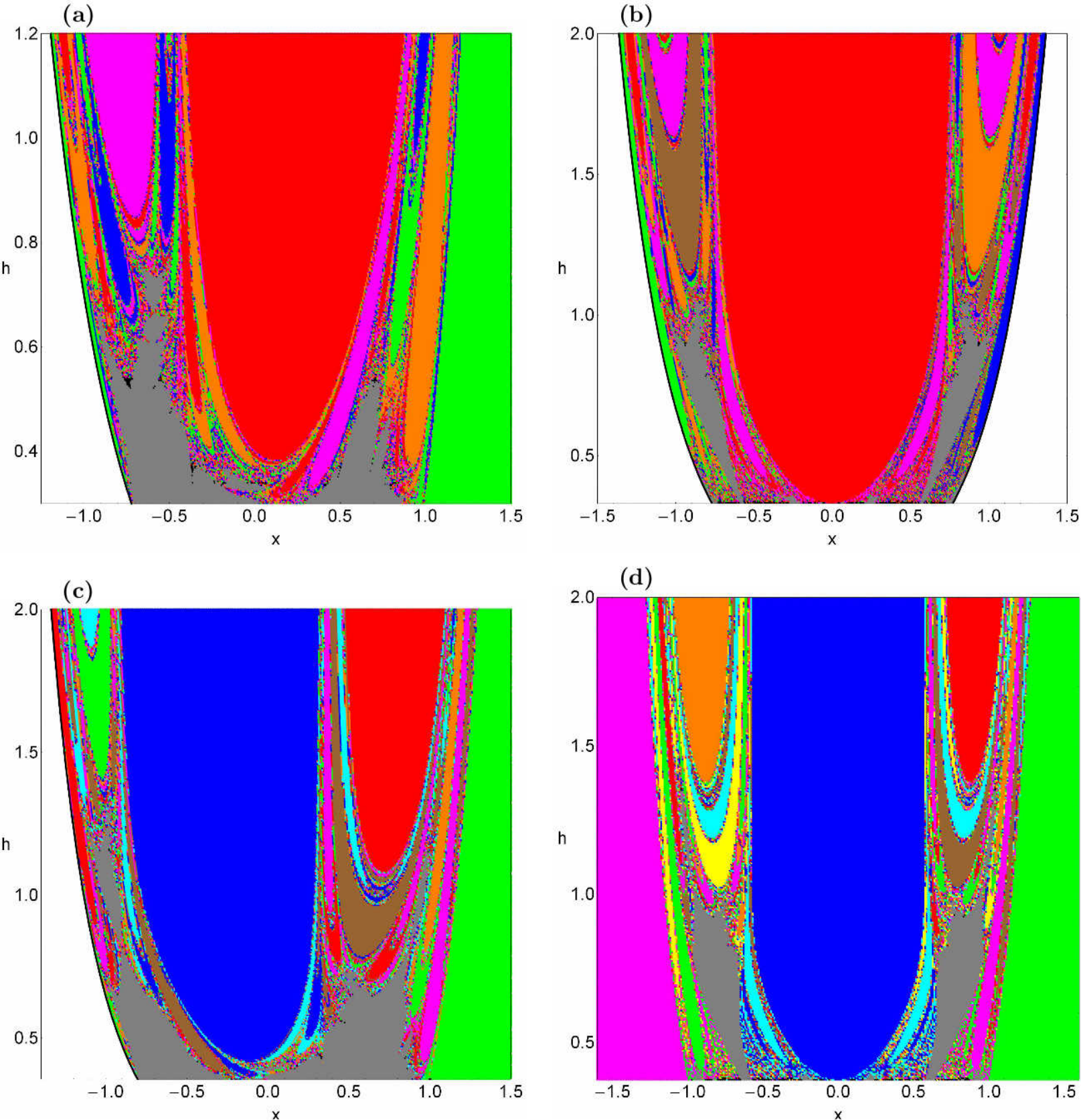}}
\caption{Orbital structure of the $(x,h)$-plane when (a-upper left): five channels of escape are present and $h \in [0.301, 1.2]$; (b-upper right): six channels of escape are present and $h \in [0.334, 2]$; (c-lower left): seven channels of escape are present and $h \in [0.356, 1,2]$; (d-lower right): eight channels of escape are present and $h \in [0.376, 2]$. These diagrams provide a detailed analysis of the evolution of the trapped and escaping orbits of the Hamiltonians when the parameter $h$ changes. The color code is the same as in Fig. \ref{xy8}.}
\label{xh}
\end{figure*}

\begin{figure*}[!tH]
\centering
\resizebox{\hsize}{!}{\includegraphics{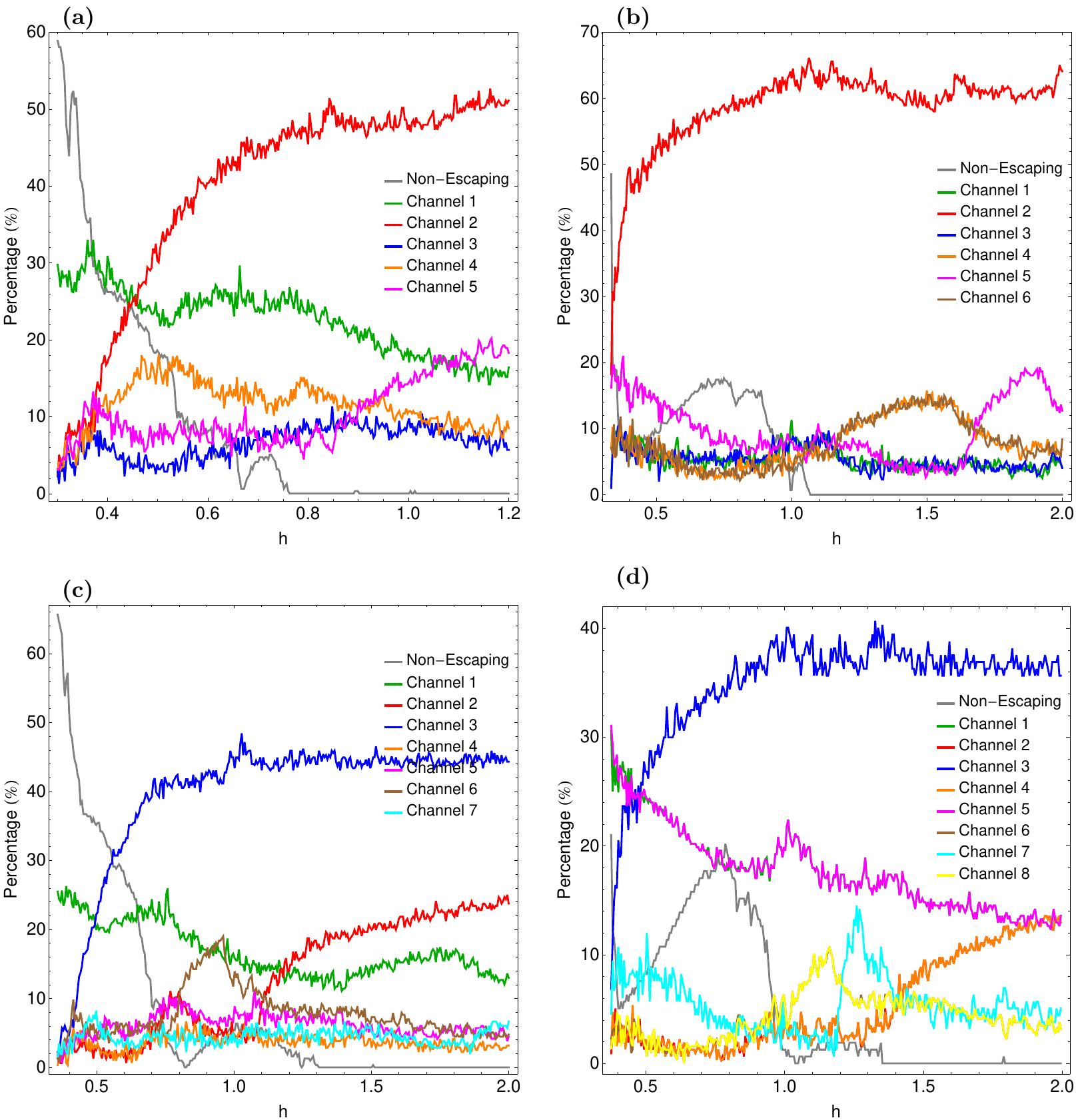}}
\caption{Evolution of the percentages of all types of orbits in the $(x,h)$ plane. (a-upper left): case with five escape channels; (b-upper right): case with six escape channels; (c-lower left): case with seven escape channels; (d-lower right): case with eight escape channels.}
\label{percs}
\end{figure*}

\begin{figure}
\includegraphics[width=\hsize]{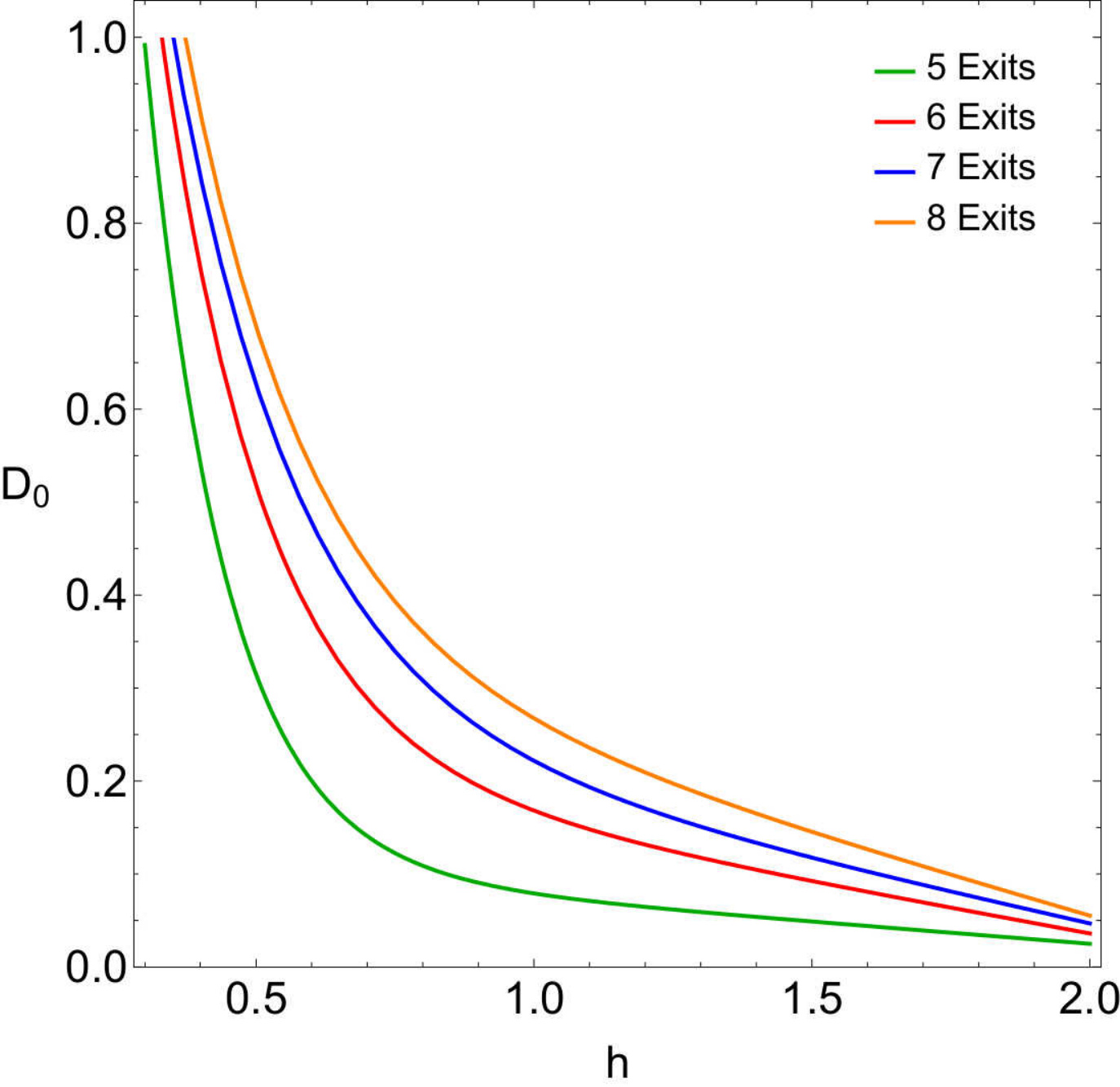}
\caption{Evolution of the fractal uncertainty dimension $D_0$ of the $(x,h)$-planes of Figs. 25(a-d) as a function of the total energy $h$. $D_0 = 1$ means total fractality, while $D_0 = 0$ implies zero fractality.}
\label{frac}
\end{figure}

\begin{figure*}[!tH]
\centering
\resizebox{\hsize}{!}{\includegraphics{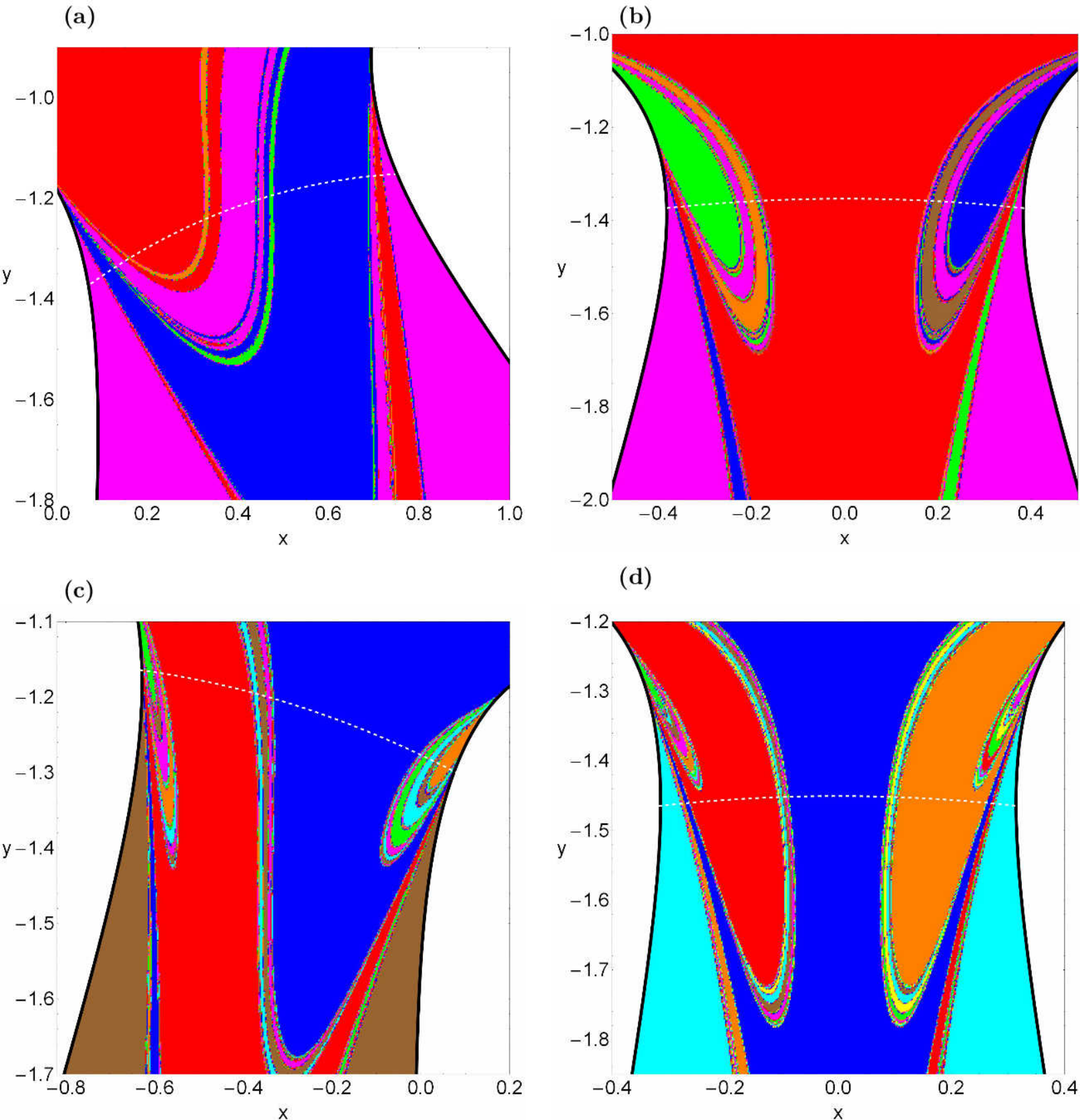}}
\caption{A zoom of the (a-upper left): exit channel 5 for $h = 0.70$; (b-upper right): exit channel 5 for $h = 1.10$; (c-lower left): exit channel 6 for $h = 1.20$; (d-lower right): exit channel 7 for $h = 1.50$;. The unstable Lyapunov orbits are shown in dashed white color.}
\label{wada}
\end{figure*}

\begin{figure*}[!tH]
\centering
\resizebox{\hsize}{!}{\includegraphics{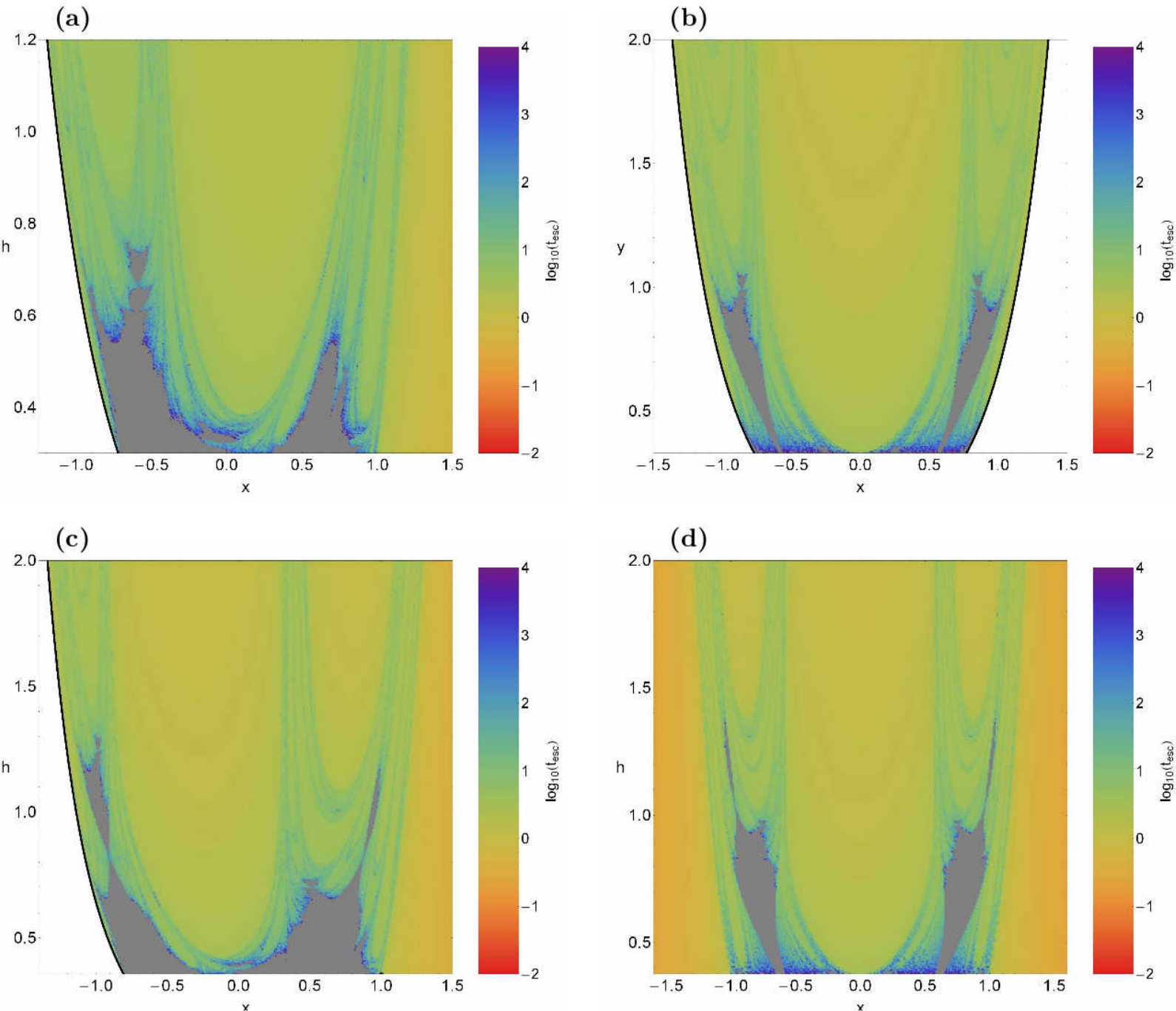}}
\caption{The distribution of the corresponding escape times of the orbits for the four types of Hamiltonians presented in Fig. \ref{xh}(a-d). In this type of grid representation the stability islands of regular orbits which are indicated by gray color can be identified more easily.}
\label{txh}
\end{figure*}

\subsection{An overview analysis}
\label{geno}

\begin{figure*}[!tH]
\centering
\resizebox{\hsize}{!}{\includegraphics{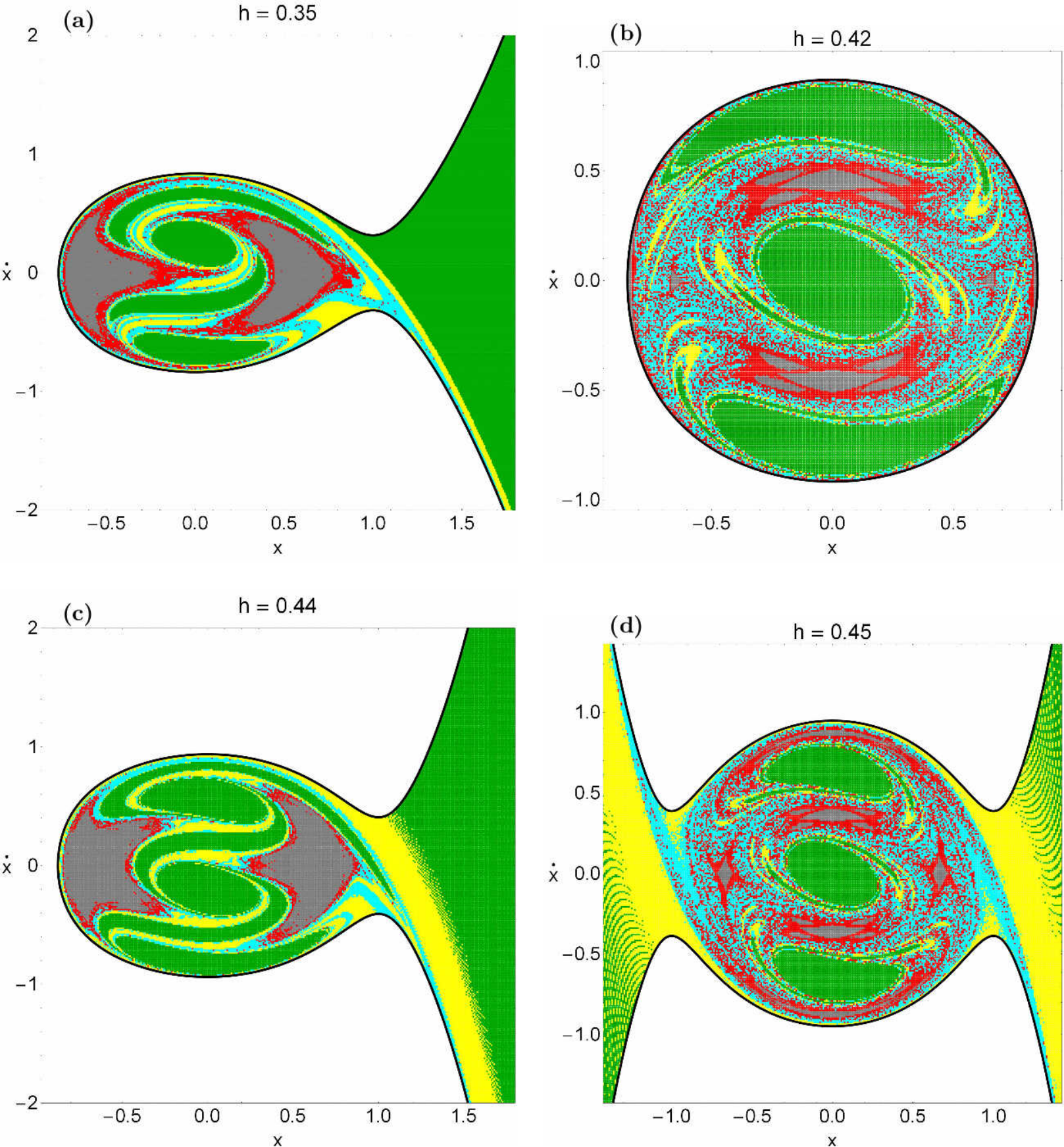}}
\caption{Color scale of the escape regions as a function of the number of intersections with the $y = 0$ axis upwards $(\dot{y} > 0)$. The color code is as follows: 0 intersections (green); 1 intersection (yellow); 2--10 intersections (cyan); $> 10$ intersections (red). The gray regions represent stability islands of trapped orbits.}
\label{iters}
\end{figure*}

\begin{figure*}[!tH]
\centering
\resizebox{\hsize}{!}{\includegraphics{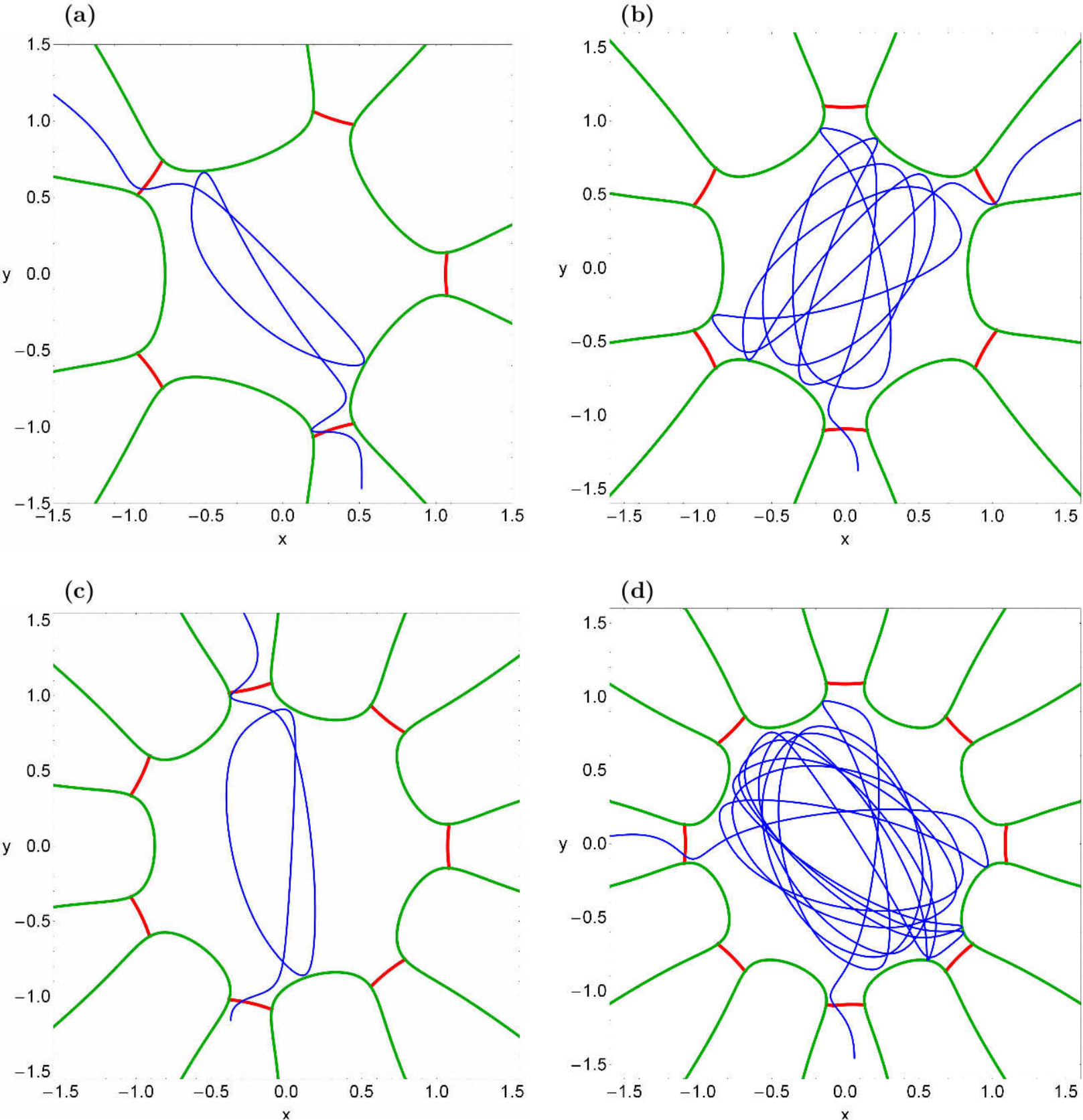}}
\caption{Characteristic examples of orbits with initial conditions outside the unstable Lyapunov orbits which however do not escape immediately from the system.}
\label{orbs}
\end{figure*}

The color-coded grids in configuration $(x,y)$ as well as the phase $(x,\dot{x})$ plane provide information on the phase space mixing however, for only a fixed value of energy. H\'{e}non back in the late 60s [\citealp{H69}], introduced a new type of plane which can provide information not only about stability and chaotic regions but also about areas of trapped and escaping orbits using the section $y = \dot{x} = 0$, $\dot{y} > 0$ (see also [\citealp{BBS08}]). In other words, all the orbits of the test particles are launched from the $x$-axis with $x = x_0$, parallel to the $y$-axis $(y = 0)$. Consequently, in contrast to the previously discussed types of planes, only orbits with pericenters on the $x$-axis are included and therefore, the value of the energy $h$ can be used as an ordinate. In this way, we can monitor how the energy influences the overall orbital structure of our Hamiltonian system using a continuous spectrum of energy values rather than few discrete energy levels. Figs. \ref{xh}(a-d) shows the structure of the $(x,h)$-plane for the four types of Hamiltonians presented in the previous subsections. It is seen that in all four plots the boundaries between bounded and unbounded motion are now seen to be more jagged than in the previous types of grids. In addition, we found in the blow-ups of the diagrams many tiny islands of stability\footnote{From chaos theory we expect an infinite number of islands of (stable) quasi-periodic (or small scale chaotic) motion.}. We observe that for low values of the energy close to the escape energy, there is a considerable amount of trapped orbits inside stability regions surrounded by a highly fractal structure. This pattern however changes for larger energy levels, where there are no trapped orbits and the vast majority of the grids is covered by well-formed basins of escape, while fractal structure is confined only near the boundaries of the escape basins. It would be of particular interest to monitor how the total orbital energy $h$ influences the percentages of all types of orbits. The following Fig. \ref{percs}(a-d) shows the evolution of the percentages of all types of orbits identified in the $(x,h)$ planes of Figs. \ref{xh}(a-d), respectively as a function of the total orbital energy.

In all previous subsections we discussed fractality of the configuration and phase space in a qualitative way. In particular, rich and highly fractal domains are those in which we cannot predict through which exit channel the particle will escape since the particle chooses randomly an exit. On the other hand, inside the escape basins where the degree of fractality is zero the escape process of the particles is well known and predictable. At this point, we shall provide a quantitative analysis of the degree of fractality for the grids shown in Figs. \ref{xh}(a-d). In order to measure the fractality we have computed the uncertainty dimension [\citealp{O93}] for different values of the total energy. Obviously, this quantity is independent of the initial conditions used to compute it. We follow the numerical way according to [\citealp{AVS01}]. We calculate the exit for certain initial condition $(x,h)$. Then, we compute the exit for the initial conditions $(x - \epsilon, h)$ and $(x + \epsilon, h)$ for a small $\epsilon$ and if all of them coincide, then this point is labeled as ``certain''. If on the other hand they do not, it will be labeled as ``uncertain''. We repeat this procedure for different values of $\epsilon$. Then we calculate the fraction of initial conditions that lead to uncertain final states $f(\epsilon)$. There exists a power law between $f(\epsilon)$ and $\epsilon$, $f(\epsilon) \propto \epsilon^{\alpha}$, where $\alpha$ is the uncertainty exponent. The uncertainty dimension $D_0$ of the fractal set embedded in the initial conditions is obtained from the relation $D_0 = D - \alpha$, where $D$ is the dimension of the phase space. It is typical to use a fine grid of values of $x$ and $h$ to calculate the uncertainty dimension. The evolution of the uncertainty dimension $D_0$ when the energy is increased is shown in Fig. \ref{frac}(a-d) for the corresponding $(x,h)$ grids of Fig. \ref{xh}(a-d), respectively. As it has just been explained, the computation of the uncertainty dimension is done for only a ``1D slice'' of initial conditions of Figs. \ref{xh}(a-d) and for that reason $D_0 \in (0,1)$. It is remarkable that the uncertainty dimension tends to one when the energy tends to its minimum value $(E_{esc})$. This means that for that critical value, there is a total fractalization of the grid, and the chaotic set becomes ``dense'' in the limit. Consequently, in this limit there are no smooth sets of initial conditions and the only defined structures that can be recognized are the Kolmogorov-Arnold-Moser (KAM)-tori of quasi-periodic orbits. When the energy is increased however, the different smooth sets appear and tend to grow, while the fractal structures that coincide with the boundary between basins decrease. Finally for values of energy much greater than the escape energy the uncertainty dimension tends to zero (no fractality). Furthermore, it is seen that there is a hierarchy in four curves shown in Fig. \ref{frac}. In particular, the order of the curves follow the number of exits (channels); the more the exits the higher the corresponding curve with more fractality. This makes sense, because if there are more basins it seems to be more probable that your closest point in the exit basin belongs to a different basin.

The rich fractal structure of the $(x,h)$ planes shown in Figs. \ref{xh}(a-d) implies that all four Hamiltonians have also a strong topological property, which is known as the Wada property [\citealp{AVS01}]. The Wada property is a general feature of two-dimensional (2D) Hamiltonians with three or more escape channels. A basin of escape verifies the property of Wada if any initial condition that is on the boundary of one basin is also simultaneously on the boundary of three or even more escape basins (e.g., [\citealp{BSBS12}, \citealp{KY91}]). In other words, every open neighborhood of a point $x$ belonging to a Wada basin boundary has a nonempty intersection with at least three different basins. Hence, if the initial conditions of a particle are in the vicinity of the Wada basin boundary, we will not be able to be sure by which one of the three exits the orbit will escape to infinity. Therefore, if a Hamiltonian system has this property the unpredictability is even stronger than if it only had fractal basin boundaries. If an orbit starts close to any point in the boundary, it will not be possible to predict its future behavior, as its initial conditions could belong to any of the other escape basins. In Fig. \ref{wada}(a-d) we present zoom plots of characteristic exit channels in the configuration $(x,y)$ space for the system with five, six, seven and eight escape channels, respectively, while the corresponding Lyapunov orbits are shown in dashed white. We see than no matter the scale, all colors are fully mixed and therefore we have an indication that our Hamiltonian system verify this special property. However, it should be pointed out that the only mathematically precise method to verify the Wada property in a Hamiltonian system is to paint the unstable manifold of the Lyapunov orbit and show that it crosses all basins (see e.g., [\citealp{NY96}]). This special topological property has been identified and studied in several dynamical systems (e.g., [\citealp{AVS09}, \citealp{KY91}, \citealp{PCOG96}]) and it is a typical property in open Hamiltonian systems with three or more escape channels.

\begin{table}
\setlength{\tabcolsep}{3.5pt}
   \caption{Initial conditions, escape period and value of the energy of the orbits shown in Fig. \ref{orbs}(a-d).}
   \label{table1}
   \begin{tabular}{@{}lcccccr}
      \hline
      Figure & $x_0$ & $y_0$ & $t_{esc}$ & $h$ & Outside & exit \\
      \hline
      \ref{orbs}a &  0.51600000 & -1.40000000 & 14.52 & 0.35 & $L_5$ & 3 \\
      \ref{orbs}b &  0.08555443 & -1.37457465 & 40.66 & 0.40 & $L_5$ & 1 \\
      \ref{orbs}c & -0.37030768 & -1.15722900 & 12.15 & 0.44 & $L_6$ & 3 \\
      \ref{orbs}d &  0.06095617 & -1.45418326 & 64.62 & 0.45 & $L_7$ & 5 \\
      \hline
   \end{tabular}
\end{table}

It is evident from the results presented in Figs. \ref{txh}(a-d) that the escape times of the orbits are strongly correlated to the escape basins. In addition, one may conclude that the smallest escape periods correspond to orbits with initial conditions inside the escape basins, while orbits initiated in the fractal regions of the planes have the highest escape rates. In all four cases the escape times of orbits are significantly reduced with increasing energy. Thus, combining all the numerical outcomes presented in Figs. \ref{xh} and \ref{txh} we may say that the key factor that determines and controls the escape times of the orbits is the value of the orbital energy (the higher the energy level the shorter the escape rates), while the fractality of the basin boundaries varies strongly both as a function of the energy and of the spatial variable. Another interesting way of measuring the escape rate of an orbit in the phase $(x,\dot{x})$ space is by counting how many intersections the orbit has with the axis $y = 0$ before it escapes. The regions in Figs. \ref{iters}(a-b) are colored according to the number of intersections with the axis $y = 0$ upwards $(\dot{y} > 0)$ and this is another type of grid representation showing a characteristic example of each Hamiltonian system. We observe that orbits with initial conditions inside the green basins escape directly without any intersection with the $y = 0$ axis. We should also note here that orbits with initial conditions located at the vicinity of the stability islands or at the boundaries of the escape basins perform numerous intersections with the $y = 0$ axis before they eventually escape to infinity. On the other hand, orbits with initial conditions inside the elongated spiral bands need only a couple of intersection until they escape.

Before closing this section, we would like to emphasize that orbits with initial conditions outside the unstable Lyapunov orbits do not necessarily escape immediately from the dynamical system. In Figs. \ref{orbs}(a-d) we present one characteristic example for each Hamiltonian and in Table \ref{table1} we provide the exact initial conditions, the escape period and the value of the energy for all the depicted orbits. We observe that even though all orbit are initiated outside but relatively close to one of the unstable Lyapunov orbits that bridge the escape channels they do not escape right away from the system. On the other hand, they enter the interior region and only after some non-zero time units of chaotic motion they eventually escape from one of the exit channels. Moreover, another interesting fact is that all four orbits escape from channels which do not coincide with the original at which they have been initiated. Thus it is evident that the initial position itself does not furnish a sufficient condition for escape, since the escape criterion is in fact a combination of the coordinates and the velocity of the test particles. More computational details regarding the escape criteria can be found in Appendix B.

\section{Conclusions and discussion}
\label{disc}

The aim of this work was to numerically investigate the escape dynamics in open Hamiltonian systems with multiple exit channels of escape. This type of dynamical systems has the key feature of having a finite energy of escape. In particular, for energies smaller than the escape value, the equipotential surfaces are closed and therefore escape is impossible. For energy levels larger than the escape energy however, the equipotential surfaces open and several channels of escape appear through which the test particles are free to escape to infinity. Here we should emphasize that if a test particle has energy larger than the escape value, this does not necessarily mean that the test particle will certainly escape from the system and even if escape does occur, the time required for an orbit to cross an unstable Lyapunov orbit and hence escape to infinity may be very long compared with the natural crossing time. The non-integrable part of the Hamiltonian containing the perturbation terms affects significantly the structure of the equipotential surface and determines the exact number of the escape channels in the configuration space. In Part I, we chose such perturbing terms creating between two and four escape channels, while here in Part II the escape channels in the $(x,y)$ plane vary between five and eight. Here we would like to emphasize that in this paper we introduce and explore for the first time potential functions that correspond to Hamiltonian systems with more than four escape channels and this is the main novelty of our work.

We defined for several values of the total orbital energy dense, uniform grids of initial conditions regularly distributed in the area allowed by the corresponding value of the energy in both the configuration $(x,y)$ and the phase $(x,\dot{x})$ space. In both cases, the density of the grids was controlled in such a way that always there were about 50000 orbits to be examined. For the numerical integration of the orbits in each grid, we needed roughly between 1 minute and 6 days of CPU time on a Pentium Dual-Core 2.2 GHz PC, depending both on the amount of trapped orbits and on the escape rates of orbits in each case. For each initial condition, the maximum time of the numerical integration was set to be equal to $10^5$ time units however, when a test particle escapes the numerical integration is effectively ended and proceeds to the next initial condition.

By conducting a thorough and systematical numerical investigation we successfully revealed the structure of both the configuration and the phase space. In particular, we managed to distinguish between trapped (non-escaping) and escaping orbits and we located the basins of escape leading to different exit channels, also finding correlations with the corresponding escape times of the orbits. Among the escaping orbits, we separated between those escaping fast or late from the system. Our extensive numerical calculations strongly suggest that the overall escape process is very dependent on the value of the total orbital energy. The main numerical results of our investigation can be summarized as follows:
\begin{enumerate}
 \item In all four Hamiltonian systems studied, areas of non-escaping orbits and regions of initial conditions leading to escape in a given direction (basins of escape), were found to exist in both the configuration and the phase space. The several escape basins are very intricately interwoven and they appear either as well-defined broad regions or thin elongated spiral bands. Regions of trapped orbits first and foremost correspond to stability islands of regular orbits where a third adelphic integral of motion is present.
 \item We observed that in several exit regions the escape process is highly sensitive dependent on the initial conditions, which means that a minor change in the initial conditions of an orbit leads the test particle to escape through another exit channel. These regions are the opposite of the escape basins, are completely intertwined with respect to each other (fractal structure) and are mainly located in the vicinity of stability islands. This sensitivity towards slight changes in the initial conditions in the fractal regions implies that it is impossible to predict through which exit the particle will escape.
 \item A strong correlation between the extent of the basins of escape and the value of the total orbital energy $h$ was found to exists. Indeed, for low values of $h$ the structure of both the configuration and the phase space exhibits a large degree of fractalization and therefore the majority of orbits escape choosing randomly escape channels. As the value of $h$ increases however, the structure becomes less and less fractal and several basins of escape emerge. The extent of these basins of escape is more prominent at relatively high energy levels, where they occupy about nine tenths of the entire area on the girds.
 \item Our numerical computations revealed that the escape times of orbits are directly linked to the basins of escape. In particular, inside the basins of escape as well as relatively away from the fractal domains, the shortest escape rates of the orbits had been measured. On the other hand, the longest escape periods correspond to initial conditions of orbits in the vicinity of stability islands or inside the fractal structures. It was also found that as we proceed to high energy levels far above the escape energy the proportion of fast escaping orbits increases significantly. This phenomenon can be justified, if we take into account that with increasing energy the exit channels on the equipotential surfaces become more and more wide thus the test particles can find easily and faster one of the exits and escape to infinity.
 \item We provided numerical evidence that our open Hamiltonian systems have a strong topological property, known as the Wada property. This means that any initial condition that is on the boundary of an escape basin, is also simultaneously on the boundary of at leats other two basins of escape. We also concluded that if a dynamical system verifies the property of Wada, the unpredictability is even stronger than if it only had fractal basin boundaries.
 \item In all four examined cases, we identified a small portion of chaotic orbits with initial conditions close enough to the outermost KAM islands which remain trapped in the neighbourhood of these islands for vast time intervals having sticky periods which correspond to hundreds of thousands time units. It should be pointed out however, that the amount of these trapped chaotic orbits is significantly smaller that that reported in the case of four exit channels of Part I.
 \item In both the configuration as well the phase space we reported the existence of streams of initial conditions which correspond to orbits that start outside the unstable Lyapunov orbits then they enter the interior region and finally escape from some escape channel which however do not coincide with the original one in which they have been initiated. These streams flow from the inside to the outside of the equipotential surfaces and extend asymptotically to infinity.
\end{enumerate}

We hope that the present numerical analysis to be useful in the active field of open Hamiltonian systems which may have implications in different aspects of chaotic scattering with applications in several areas of physics. For example, we related the current model potential with applications in the field of reactive multichannel scattering. Moreover, it is in our future plans to expand our investigation in other more complicated potentials, focusing our interest in reveling the escape mechanism of stars in galactic systems such as star clusters, binary stellar systems, or barred spiral galaxies.

\section*{Acknowledgments}

I would like to express my warmest thanks to Prof. James D. Meiss and Jacobo Aguirre for all the illuminating and inspiring discussions during this research and also to Prof. Christof Jung for pointing out the interesting subject of multichannel chaotic scattering. My thanks also go to the two anonymous referees for the careful reading of the manuscript and for all the apt suggestions and comments which allowed us to improve both the quality and the clarity of the paper.

\section*{Compliance with Ethical Standards}

\begin{itemize}
  \item Funding: The author states that he has not received any research grants.
  \item Conflict of interest: The author declares that he has no conflict of interest.
\end{itemize}

\section*{APPENDIX A: LIST OF PERTURBATION FUNCTIONS}
\label{apex1}

In the following Table \ref{table2} we provide the equations containing the perturbing terms derived by the generating functions (\ref{gens}), for the first nine cases, that is when the Hamiltonian system has between two and ten channels of escape in the configuration $(x,y)$ space. Note that in Part I for the case of four exits we adopted the perturbation function $V_1(x,y) = - x^2 y^2$, simply because it was also used in many earlier works, while in Table (\ref{table2}) we give the general function according to the corresponding generating function.

\begin{table}[!ht]
\setlength{\tabcolsep}{3pt}
   \caption{Equations of perturbing terms when $n \in [2, 10]$.}
   \label{table2}
   \begin{tabular}{@{}lr}
      \hline
      Channels & Perturbation function $V_1(x,y)$\\
      \hline
       $n = 2$ & $V_1 =  - x y^2$ \\
       $n = 3$ & $V_1 =  - \frac{1}{3} (x^3 - 3 x y^2)$ \\
       $n = 4$ & $V_1 =  - \frac{1}{4} (x^4 - 6 x^2 y^2 + y^4)$ \\
       $n = 5$ & $V_1 =  - \frac{1}{5} (x^5 - 10 x^3 y^2 + 5 x y^4)$ \\
       $n = 6$ & $V_1 =  - \frac{1}{6} (x^6 + 15 x^4 y^2 - 15 x^2 y^4 + y^6)$ \\
       $n = 7$ & $V_1 =  - \frac{1}{7} (x^7 - 21 x^5 y^2 + 35 x^3 y^4 - 7 x y^4)$ \\
       $n = 8$ & $V_1 =  - \frac{1}{8} (x^8 - 28 x^6 y^2 + 70 x^4 y^4 - 28 x^2 y^6 + y^8)$ \\
       $n = 9$ & $V_1 =  - \frac{1}{9} (x^9 - 36 x^7 y^2 + 126 x^5 y^4 - 84 x^3 y^6 + 9 x y^8)$ \\
      $n = 10$ & $V_1 = - \frac{1}{10} (x^{10} - 45 x^8 y^2 + 210 x^6 y^4 - 210 x^4 y^6 + 45 x^2 y^8 - y^{10})$ \\
      \hline
   \end{tabular}
\end{table}

\section*{APPENDIX B: ESCAPE PROCEDURE \& CRITERIA}
\label{apex2}

\begin{figure}
\includegraphics[width=\hsize]{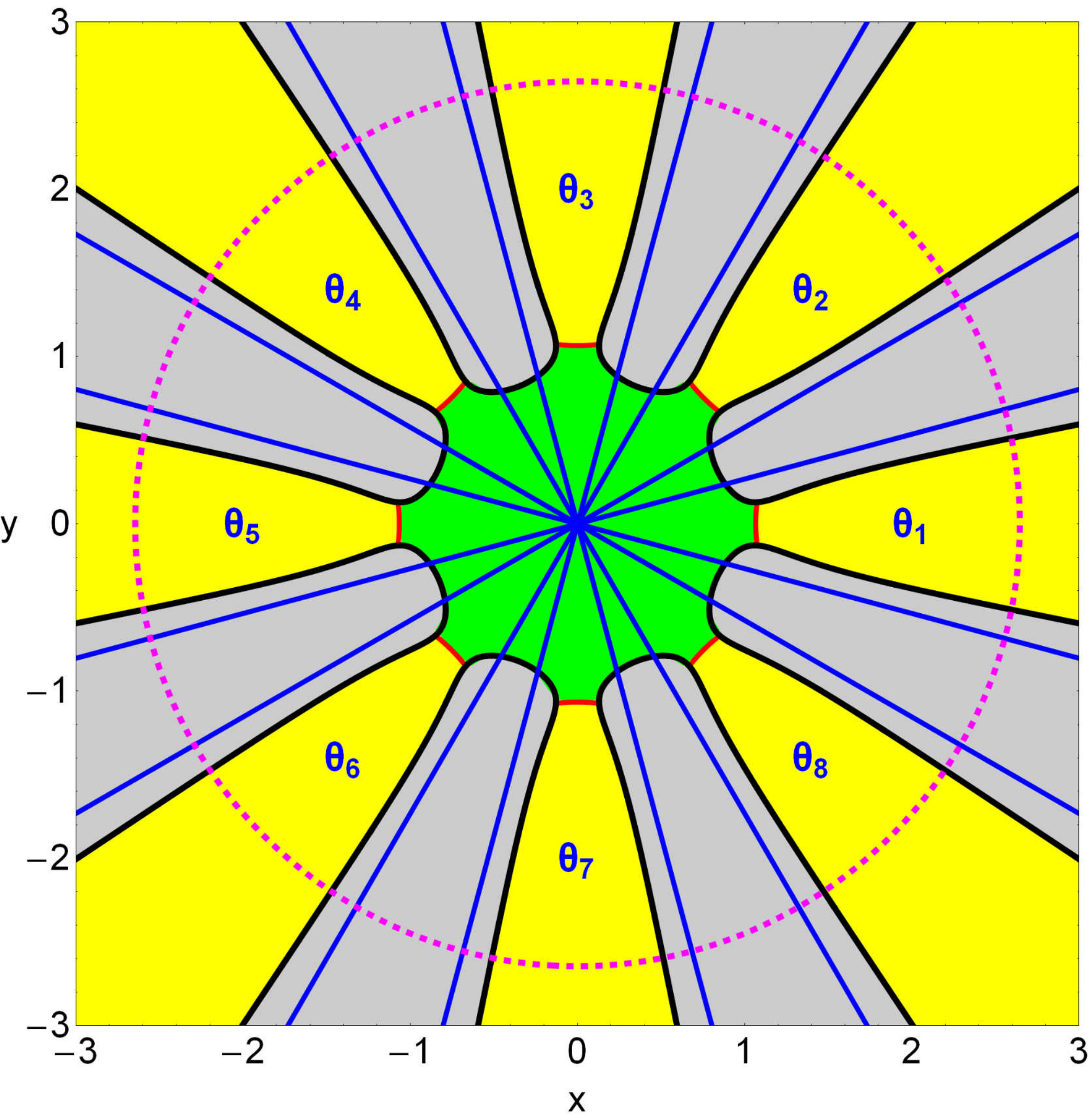}
\caption{The equipotential curve for the Hamiltonian with eight channels of escape when $h = 0.45$ is shown in black color, while the unstable Lyapunov orbits are indicated with red color. The configuration plane is divided into three domains: (i) the interior region (green), the exterior region (yellow) and (iii) the forbidden regions (gray). The blue straight lines define the angular sectors for each channel of escape, while the dashed, magenta line corresponds to the limiting circle.}
\label{ang}
\end{figure}

Here we would like to present a step by step explanation of the escape procedure of orbits and analyze all the corresponding computational aspects. We consider the case of the Hamiltonian system with eight channels of escape (obviously in all other cases with less escape channels things are much simpler) and we choose the energy level $h = 0.45 > h_{esc}$. In Fig. \ref{ang} the corresponding equipotential curve is shown in black, while the eight unstable Lyapunov orbits are denoted using red color. The initial conditions $(x_0,y_0)$ of orbits in the configuration space are divided into two main categories: (i) orbits with initial conditions in the interior region (green), that is inside the Lyapunov orbits and (ii) orbits initiated at the exterior region (yellow), that is outside the Lyapunov orbits. The gray regions on the other hand, correspond to the forbidden area where motion is impossible.

Let us first deal with the orbits initiated in the interior region. It is evident from Fig. \ref{ang} that the escape channels are very close to one another and this behavior becomes stronger in Hamiltonians with more exits $(n > 8)$. However, in any case, it is possible to define appropriate angles that embrace each channel as it is seen in Fig. \ref{ang}. Due to the overall symmetry of the dynamical system it is $\theta_1 = \theta_2 = \theta_3 = \theta_4 = \theta_5 = \theta_6 = \theta_7 = \theta_8 = 30^{\circ}$. Now we need to determine where each angle starts and where it ends so as to divide the configuration space into eight angular sectors. For this purpose, we define a polar angle which starts counting from the $x$-axis $(y = 0)$. Then we have for each sector

sector 1: $\theta_1 < 15$ or $\theta_1 > 345$,

sector 2:  $30 < \theta_2 < 60$,

sector 3:  $75 < \theta_3 < 105$,

sector 4: $120 < \theta_4 < 150$,

sector 5: $165 < \theta_5 < 195$,

sector 6: $210 < \theta_6 < 240$,

sector 7: $255 < \theta_7 < 285$,

sector 8: $300 < \theta_8 < 330$.

Along each time step of the numerical integration we monitor the position of the test particle given by the coordinates $(x,y)$ as well its velocity vector. When a test particle crosses one of the Lyapunov orbits with velocity pointing outwards then the escape takes place. In order to determine through which exit channel, or in other words through which sector the orbit has escaped we need to calculate the corresponding angle through the Cartesian coordinates. Therefore we define $z = y/x$ and the polar angle reads
\begin{equation}
\theta = \left\{
\begin{array}{lr}
  \tan^{-1}(z),        &\mbox{  if $x > 0$ and $y \geq 0$,} \\
  \tan^{-1}(z) + \pi,  &\mbox{  if $x < 0$,} \\
  \tan^{-1}(z) + 2\pi, &\mbox{  if $x > 0$ and $y < 0$,}
\end{array} \right.
\label{angz}
\end{equation}
where the output is given in radians. We can easily transform the result into degrees by multiplying with $180^{\circ} \theta/\pi$. Thus following this procedure we can determine the exit channels of orbits initiated in the interior region.

Orbits with initial conditions outside the unstable Lyapunov orbits exhibit a different behavior. In Fig. \ref{orbs}(a-d) we saw that orbits with initial conditions in the exterior region do not escape directly to infinity but on the other hand they enter the interior region and after some countable (non-zero) time they escape. For this type of orbits we use the above-mentioned technique for determining the exact channel of escape. However, the vast majority of orbits with initial conditions in the exterior region escape directly to infinity without entering the interior region and therefore crossing any Lyapunov orbit. In this case, we consider an orbit to escape when $x^2 + y^2 > q$, where $q$ is a real number depending in the particular dynamical system (for $n = 8$ we have $q = 7$). We may say that the equality $x^2 + y^2 = q$ defines a limiting circle that determines the escape of orbits initiated in the exterior region.

\begin{figure}
\includegraphics[width=\hsize]{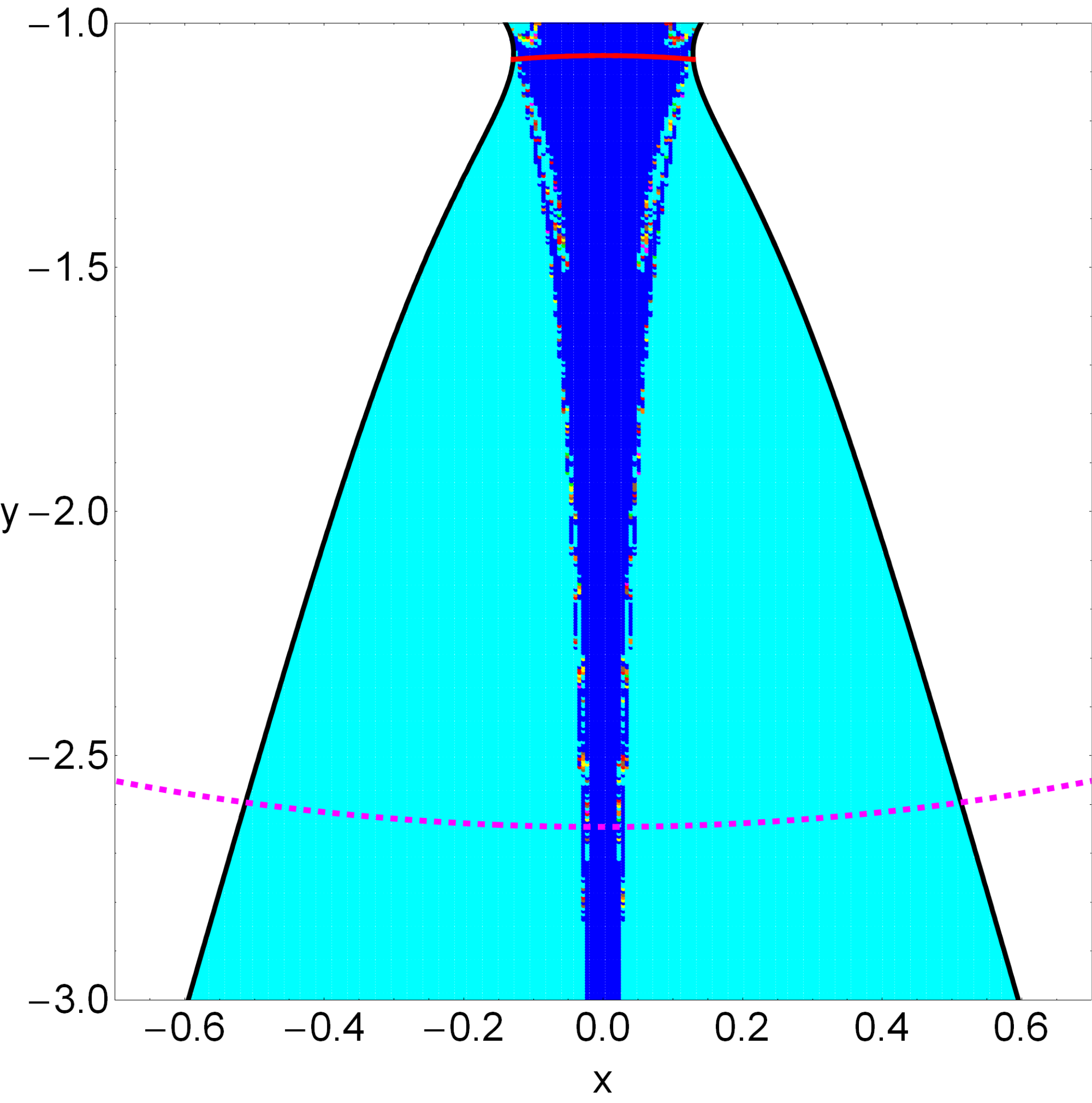}
\caption{Magnification of channel 7 when $h = 0.45$. The stream of blue initial conditions corresponding to exit 3, flows outside the unstable Lyapunov orbit (red) and extends vertically to infinity.}
\label{flow}
\end{figure}

When studying the escape dynamics of the configuration space we found the existence of streams of initial conditions which correspond to orbits that start outside the Lyapunov orbits, then they enter the interior region and finally escape from an exit which however do not coincide with the original one in which they have been initiated. Fig. \ref{flow} shows a magnification of channel 7 when $h = 0.45$. We observe the stream of blue initial conditions corresponding to exit 3, that flows outside the unstable Lyapunov orbit and extends vertically to infinity. In the same figure we plotted the limiting circle for $q = 7$. It is evident that the value of $q$ strongly depends on the size of the grid. In our calculations we considered in all four cases initial conditions of orbits inside the square area $-2 \leq x \leq 2$ and $-2 \leq y \leq 2$. For creating Fig. \ref{flow} where $y_{max} = -3$, we increased the radius of the limiting circle to $q = 12$, in order to correctly determine the escape process of orbits in the outflow stream.

\end{document}